\documentclass[twocolumn,english,prb,floatfix,superscriptaddress,twocolumn,amsmath,amssymb,reprint]{revtex4}

\usepackage[T1]{fontenc}
\usepackage[latin9]{inputenc}
\usepackage{color}
\definecolor{shadecolor}{rgb}{1, 0, 0}
\usepackage{verbatim}
\usepackage{float}
\usepackage{framed}
\usepackage{amsmath}
\usepackage{amssymb}
\usepackage{graphicx}

\makeatletter
\PassOptionsToPackage{caption=false}{subfig}
\usepackage{hyperref}
\hypersetup{
breaklinks=true,
colorlinks=true,
citecolor=blue,
linkcolor=blue,
filecolor=blue,
urlcolor=blue
}
\IfFileExists{lmodern.sty}{\usepackage{lmodern}}{}
\usepackage{babel}

\begin{document}

\title{Anyons in Multichannel Kondo Systems}

\author{Pedro L. S. Lopes}
\affiliation{Department of Physics and Stewart Blusson Institute for Quantum Matter, University of British Columbia, Vancouver, Canada V6T 1Z1}



\author{I. Affleck}
\affiliation{Department of Physics and Stewart Blusson Institute for Quantum Matter, University of British Columbia, Vancouver, Canada V6T 1Z1}

\author{E. Sela}
\affiliation{Department of Physics and Stewart Blusson Institute for Quantum Matter, University of British Columbia, Vancouver, Canada V6T 1Z1}
\affiliation{Raymond and Beverly Sackler School of Physics and Astronomy, Tel-Aviv University, IL-69978 Tel Aviv, Israel}

\begin{abstract}
    Fractionalized quasiparticles - anyons - bear a special role in present-day physics. At the same time, they display properties of interest both foundational, with quantum numbers that transcend the spin-statistics laws, and applied, providing a cornerstone for decoherence-free quantum computation. The development of platforms for realization and manipulation of these objects, however, remains a challenge. Typically these entail the zero-temperature ground-state of incompressible, gapped fluids. Here, we establish a strikingly different approach: the development and probing of anyon physics in a gapless fluid. The platform of choice is a chiral, multichannel, multi-impurity realization of the Kondo effect. We discuss how, in the proper limit, anyons appear at magnetic impurities, protected by an asymptotic decoupling from the fluid and by the emerging Kondo length scale. We discuss possible experimental realization schemes using integer quantum Hall edges. The gapless and charged degrees of freedom coexistent with the anyons suggest the possibility of extracting quantum information data by transport and simple correlation functions. To show that this is the case, we generalize the fusion ansatz of Cardy's boundary conformal field theory, now in the presence of multiple localized perturbations. The generalized fusion ansatz captures the idea that multiple impurities share quantum information non-locally, in a way formally identical to anyonic zero modes. We display several examples supporting and illustrating this generalization and the extraction of quantum information data out of two-point correlation functions. With the recent advances in mesoscopic realizations of multichannel Kondo devices, our results imply that exotic anyon physics may be closer to reach than presently imagined.
\end{abstract}

\maketitle

\section{Introduction}

The search for anyons, and in particular their richer subclass of non-Abelian anyons, comprises a multipurpose quest in present-day physics.~\cite{moore_read,STERN2008204,RevModPhys.80.1083} From a fundamental aspect, these particles carry quantum numbers that defy the concept of fundamental particles of isolated constituents of matter. They furthermore refuse to obey the standard theorems of spin-statistics; the study of their classification and exchange properties motivated extensions of group theory into fusion and braiding category theories~\cite{KITAEV_honey} (see Refs.~\onlinecite{pachos2012introduction,BerneNeupert_lectures} for digestible reviews). Allowance of this is due to a reduced dimensionality: anyons typically exist in two spatial dimensions. These properties result in arguably their most remarkable feature, namely that a set of (non-Abelian) anyons describe a Hilbert space which can be explored by in-plane exchange manipulations, braidings. Its state information is stored non-locally by the whole set of spatially distributed anyons.

Jointly the above qualities culminate in a promising practical aspect of the quest, that of performing decoherence-free topological quantum computation.~\cite{KITAEV_honey,Nayak_review} The principal bottleneck for the realization of quantum computation hardware regards suppressing decoherence. \emph{Local} perturbations destroy the quantum entanglement needed to perform complex calculations. One solution is performing error correction.~\cite{Kitaev_error_cor,Devitt_error_cor} In an oversimplified explanation, one creates redundancies by grouping sets of physical qubits in logical ones. Errors in the former are less impacting in the calculations of the latter. However, error correction costs a huge overhead to achieve fault-tolerance. Estimates of particular prime-decomposition algorithms point to ratios of order $\sim10^4$ between physical and logical qubits.~\cite{PhysRevA.86.032324} The state-of-the-art 
quantum hardware relies on superconducting devices, which remain still in the scale of $\sim 10$s of qubits.~\cite{PhysRevLett.119.180511,20qubits} By virtue of the \emph{non-locality} of the quantum information stored and manipulated with non-Abelian anyons, the decoherence problem is avoided at its most fundamental level.

Pursuing and manipulating non-Abelian anyons thus presents a fundamental, and potentially useful, field of research. Somewhat awkwardly, however, the main challenge of this search remains a most basic one: what is the ideal breeding ground for non-Abelian particles? Historically, developments focused on 2D topological phases, incompressible (gapped) fluids with, typically, protected edge modes.~\cite{moore_read} The paradigmatic anyon system involves specifically the composite particles of the fractional quantum Hall effect (FQHE).~\cite{STERN2008204,FQH_anyons_review} Nevertheless, the general focus on gapped systems led to a prolific development in theory and experiments on topological order way beyond the FQHE. This includes spin liquids~\cite{KITAEV_honey,bauer2014chiral,nasu2015thermodynamics,gorohovsky2015chiral,meng2015coupled,yao2007exact}, topological superconductivity and symmetry protected phases,~\cite{fu2008superconducting} as well as topological phases nanowires~\cite{lutchyn2010majorana,oreg2010helical,lutchyn2018majorana} and artificially engineered low-dimensional structures.~\cite{Lu_Vish,lindner2012fractionalizing,clarke2013exotic,vaezi2013fractional,klinovaja2014time,oreg2014fractional,Stanescu_2013,Gu_Wen,Mong_coupled,Sahoo_coupled,golan2017majorana,Hu_coupled,Lopes_coupled,Iadecola_coupled_wires,burrello2013topological,landau2016towards}

The gap that protects these topological systems has been understood as compulsory for the existence of anyons. Indeed, in the 2D scenarios described above, anyons exist as dynamical excitations over an isolated ground multiplet. Anyon-pair creation, braiding, and subsequent annihilation, in the non-Abelian case, manipulates the states of this ground multiplet. These ground states must be isolated from the remainder of excited states of the system, if manipulation by adiabatic anyon exchanges are to take place at all.~\cite{RevModPhys.80.1083} From a practical perspective, however, systems where anyons exist dynamically are not ideal - anyon pairs may, in this case, still be created due to thermal effects, interfering with intended braids and leading to errors. In some sense, ``universally'' useful non-Abelian anyons must also be bound or confined to some local structure, such as surfaces, edges, or defects of order parameters, like superconducting vortices.~\cite{UFQP} If so, this dramatically reduces the number of ideal anyonic candidate-platforms, which would be summarized by topological superconductors and some cases of artificially engineered coupled-wires and other networks.

The principle of this ``gap dogma'' for anyons can be described as a splitting of the Hilbert space by an energy scale, with a ground multiplet where the anyons act (non-locally) separated from the remainder states. The main goal of this work is to expand this horizon.  Our motivating question is: can we extract a state sector from a system in which to store quantum information, non-locally, without the need of an energy gap? We show that the answer for this question is affirmative. Such an emergent splitting of a Hilbert space in orthogonal sectors is possible by relying on many-body competition effects, frustration, leading to non-Fermi-liquid (NFL) physics. The scenario we propose is based on a multi-impurity chiral version of the multi-channel Kondo effect.

\begin{figure}[tb]
	\centering
	\includegraphics[width=1.0\columnwidth]{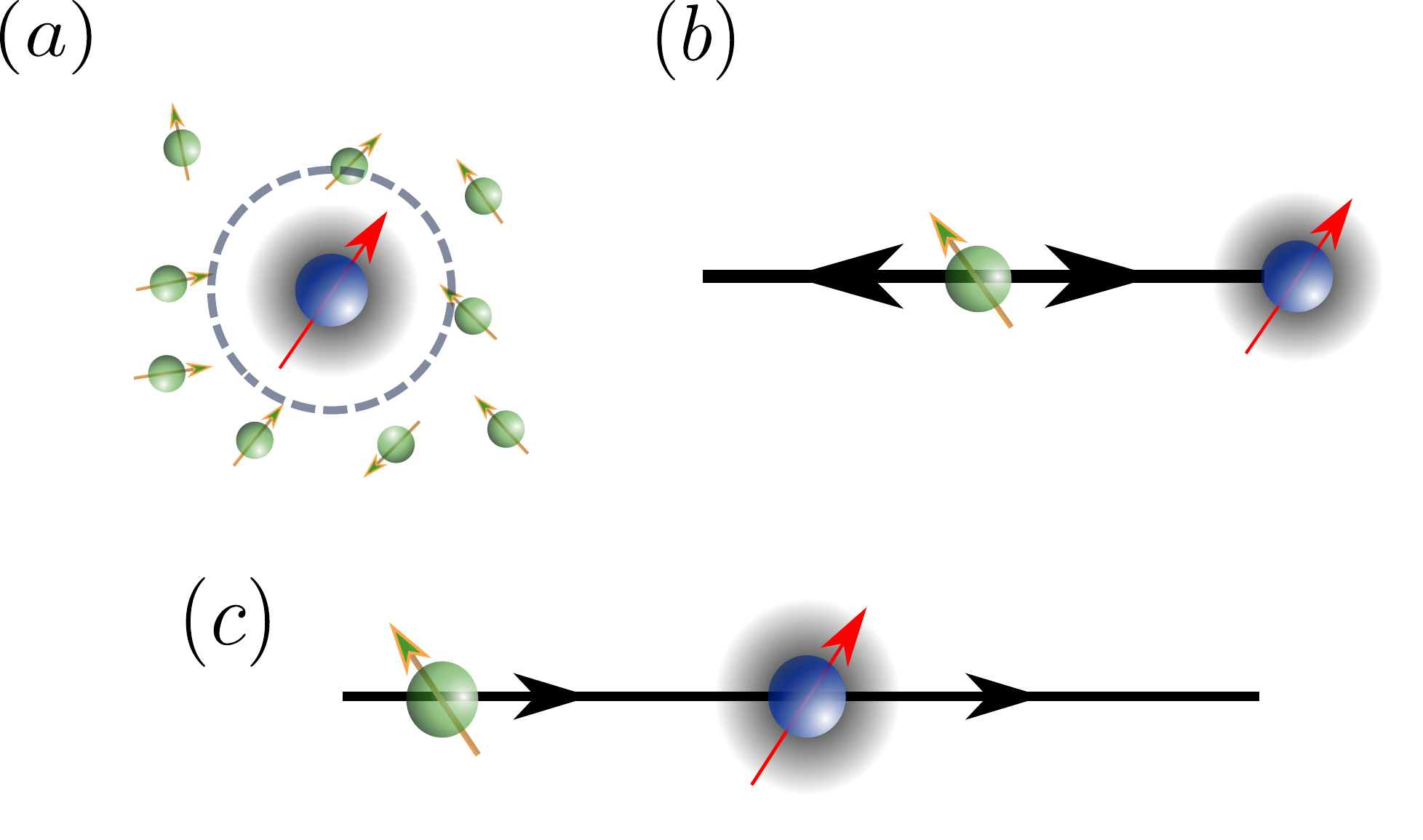}
	\caption{Three equivalent versions of the Kondo problem:
	(a) a bath of spinful electrons (green balls) interacting with a magnetic impurity (blue ball) via s-wave scattering (dashed circle)  are equivalent to (b) a 1D radial problem, where particles propagate from infinity and scatter at the origin containing the impurity. (c) This 1D picture can be ``unfolded'', resulting in a chiral version where electrons move in a single direction and scatter off the impurity at the origin.}
	\label{fig:Kondos}
\end{figure}

The Kondo effect is a paradigmatic phenomenon of strongly correlated electrons. Originally motivated by a logarithmic increase of resistivity in certain metals at low-temperatures, the phenomenon is (toy-)modelled by an electronic fluid interacting with a magnetic impurity.~\cite{Kondo,Anderson_Kondo,Schrieffer_Wolff_Kondo,hewson1997kondo} In its simplest depiction, a fluid containing a single flavor, or channel, of spinful electrons is put in contact with a spin-1/2 impurity [Fig.~\ref{fig:Kondos}(a)]. At low energies, such an impurity is absorbed by the electron fluid: the impurity magnetic moment is perfectly screened by the electrons, effectively forming a singlet with one of them. The other electrons are completely expelled from the impurity, simply becoming free. The only residual effect on these free conduction degrees of freedom is a phase shift -- due to a hard-core scattering at the screened impurity position -- and they behave as a regular Fermi liquid.~\cite{Nozieres1974} A plethora of ramifications and modifications of this scenario have been studied in the literature. These include higher-spin impurities,\cite{hewson1997kondo,Affleck_NCK} Kondo impurity lattices,\cite{varma_Kondo_lattice,Coleman_Kondo_Lattice,millislee_Kondo_lattice,Auerback_Kondo_Lattice,Rice_Ueda_Kondo_lattice,Stewart_kondo_lattice} coupled impurities,\cite{Affleck_2IK,Affleck_2IK_NRG,georges1999electronic,zarand2006quantum,mitchell2011two,mitchell2012two} and, most important to the present discussion, multi-channel versions of the problem.~\cite{nozieres1980kondo,vigman1980pis,andrei1980diagonalization,AFFLECK_Ludwig_NCK1,LUDWIG_NCK,cox1998exotic}

The $N$-channel Kondo (NCK) effect is naively a modification of the discussion above where several flavors of fermionic degrees of freedom are in exactly symmetric contact with a single impurity. This brings, however, remarkably deep consequences. For our purposes it will suffice to stick to the spin-1/2 impurity scenario. In this case, the many competing electronic channels are seen to ``overscreen'' the spin-1/2 impurity which can't choose a particular channel to form a singlet with. Frustrated, the option out  for the impurity is to fractionalize; part of its degrees of freedom are absorbed by the 
conduction electrons which then behave as a NFL. Crucially, the remnants of the fractionalized impurity are completely decoupled from the Hilbert space of the gapless liquid and, from the viewpoint of low-energy and long-wavelength probes,~\cite{affleck2001detecting,yoo2018detecting,Affleck_Kondo_cloud,Kondo_Cloud_exp} are localized around the impurity site by an emergent Kondo length scale $\xi_K$.

The onset of NFL behavior and emergent decoupling of degrees of freedom in the NCK problem has a twofold signature.~\cite{Affleck_Kondo_review} First, the correlation functions for electrons scattering off the impurity display a scattering S-matrix with modulus less than unity. This non-unitarity suggests that part of the electronic degrees of freedom, at times their totality, is disappearing from the problem after being in contact with the impurity; naturally they are being transformed into something else,\footnote{This ``something else'' can be tracked exactly in the two-channel --2CK-- case; see Ref.~\onlinecite{MALDACENA_Ludwig_SO8}.} a NFL type of behavior.~\cite{nozieres1980kondo} Second, in the thermodynamic limit $L\to \infty$, it has been shown that even at the lowest temperatures a finite entropy remains in the NCK system. This ``impurity entropy'' has the form
\begin{equation}
    S^{(N)}_{\rm{imp}}=\log g(N)
\end{equation}
which is computed in the $T\to0$ limit, taken after $L\to\infty$. $N$ is the number of channels, and~\cite{vigman1980pis,andrei1980diagonalization,Affleck_Ludwig_entropy}
\begin{equation}
    g(N)=2\cos\left(\frac{\pi}{N+2}\right)
\end{equation}
The order of limits is important.\cite{Rozhkov_imp_entropy,Affleck_Kondo_review} The system must be allowed to thermally access a continuum of states for the onset of this behavior. For a single channel, nothing is left at the impurity as $g(1)=1$, but for larger $N$ a distinction immediately appears. For example, $g(2)=\sqrt{2}$, while $g(3)=(1+\sqrt{5})/2$. How should these leftover degrees of freedom be interpreted? Connecting to topological quantum information, we remark that these numbers correspond exactly, and respectively, to the quantum dimensions of Ising and Fibonacci anyons. The quantum dimension $d_a$ of an anyon $a$ dominates the dimension of the Hilbert space of a large number of anyons of type $a$. The idea that the objects left behind at an overscreened impurity are very similar to anyons of topological phases receives strong support from the two-channel case. This problem has a famous exact solution by Emery and Kivelson in the anisotropic, so-called, Toulouse point.~\cite{Emery_Kivelson} In this case a Majorana zero-mode is explicitly shown to be decoupled from the conduction degrees of freedom at the impurity position. 

Under the hypothesis that NCK systems display localized and fractionalized particles, we remark that it presents also an ideal platform from the point of view of manipulation. For this, one relies on mesoscopic realizations of the Kondo problem. Historically, the Kondo effect originated in the context of solid-state physics and 3-dimensional materials with free s-d-shell electrons scattering.
~This scattering, however, is predominantly dominated by a fully isotropic s-partial-wave component and, in the single-impurity case, the only spatial degree of freedom of importance is the radial coordinate [Fig.~\ref{fig:Kondos}(b)]. The problem is effectively 1D and, in fact, the key ingredient is just a finite density of electronic states interacting with a local impurity. These observations allowed a subsequent extension of the Kondo effect to a bevy of scenarios, including mesoscopic circuits and quantum dots.~\cite{goldhaber1998kondo,Cronenwett_Kondo_QPC} Here, the Coulomb interaction in a gated quantum dot leads to a control over electronic tunneling into or out of the dot which, depending on the physical realization, effectively behaves as a two-level system, a pseudospin, like a magnetic impurity. The pseudospin can be composed of spin degrees of freedom if the charge of the dot is frozen due to the Coulomb blockade effect,~\cite{goldhaber1998kondo,Cronenwett_Kondo_QPC,pustilnik2004kondo} or alternatively, it can derive from the charge degree of freedom when a gate voltage takes the system close to a degeneracy between two macroscopic charge states.~\cite{furusaki1995theory} Coupling it to an electronic reservoir results in a mesoscopic realization of a charge Kondo problem. Given the speed, precision, and scaling  of present-day microelectronics, on-demand engineered mesoscopic systems with anyonic behavior are worthy contenders for quantum information applications (as attested by the recent interest in the pursuit of Majorana fermions in nanowires~\cite{lutchyn2010majorana,oreg2010helical}). The necessary perfect symmetry between channels presents, however, a challenge for the realization of the multi-channel Kondo effect, in all scenarios. Nevertheless, it has also been circumvented in the mesoscopic approach, where the Coulomb blockade has been exploited to achieve a full independence of electronic reservoirs coupled to a quantum dot.~\cite{Potok_GG_2CK,keller2015universal} Experimental signatures of the two-channel Kondo effect and, recently, even of the three-channel case are  presently available using a charge pseudospin.~\cite{Iftikhar_QPC2,Iftikhar_QPC1,Iftikhar_Thesis,Iftikhar_3CK}

\begin{figure}[tb]
	\centering
	\includegraphics[width=1.0\columnwidth]{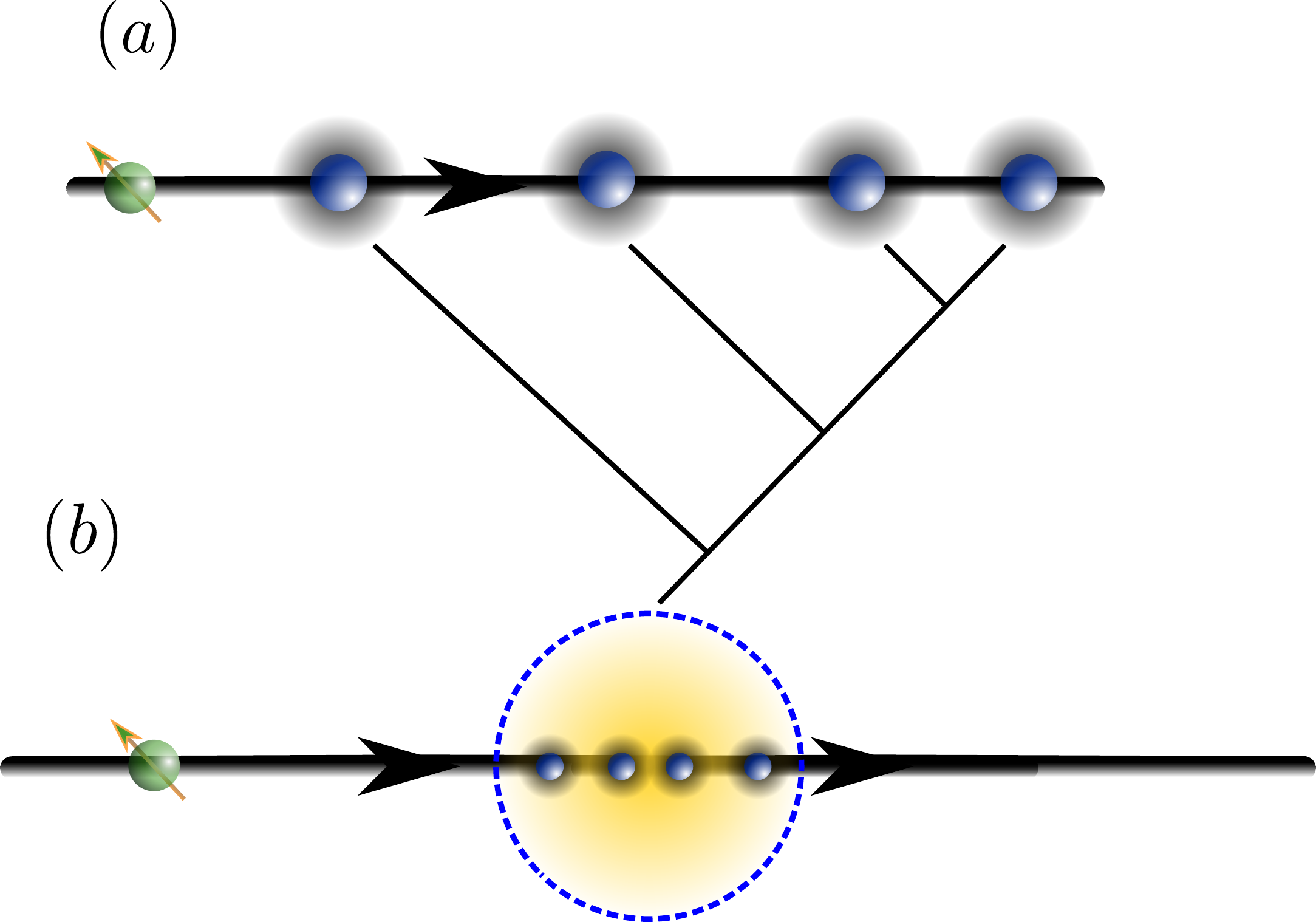}
	\caption{A chiral multi-impurity model. (a) An electron (green ball) of a single chirality moves through a set of impurities (blue balls). The 
	scattering region around each impurity is assumed to be localized, characterized by a length scale -- such as the Kondo length -- which is assumed to be smaller than the inter-impurity  separation. The impurity degrees of freedom are thus assumed to be independent of each other. (b) The impurities may be fused into a single effective perturbation (dashed blue and yellow region). This last picture 
	can again be seen in folded or unfolded pictures.}
	\label{fig:Kondo_multi}
\end{figure}

So far we have argued for the existence of quasiparticle fractionalization in NCK systems. Furthermore, we reviewed the experimental relevance and feasibility of NCK in highly tunable and accessible devices. We may finally concentrate on our main results. The main deficiency of the points made above is (i) a candidate scenario where to realize multiple ``Kondo anyons'' in a way that (ii) it is possible to extract their non-locally stored information. To address these points, we exploit an unfolded 1D picture 
of the NCK problem [Fig.~\ref{fig:Kondos}(c)] and extend it to a multi-impurity case as displayed in Fig.~\ref{fig:Kondo_multi}. We propose an implementation of this geometry via an integer quantum Hall device. Following the scenario of the impurity entropy, we discuss how correlation functions, when evaluated at positions far but surrounding the impurities, display a normalization factor that depends globally on the effects from all enclosed impurities. In the case when anyons are confined to the impurities, this normalization factor depends on their effective fusion channel, giving us access to the non-local anyonic Hilbert space.

The convenience of the 1D realization of the Kondo problem deserves discussion. It allows the implementation of conformal field theory (CFT) techniques, arguably one of the most powerful tools for dealing with strongly correlated matter, albeit available only in gapless 1D fluids.~\cite{BYB} The change in correlation functions due to the presence of a local edge impurity is captured by ``boundary conformal field theory'' (BCFT), developed in the seminal works of Cardy~\cite{CARDY_1,CARDY_2,Cardy_Lewellen} (for reviews see e.g. Refs.~\onlinecite{Cardy_review,Affleck_Kondo_review}, and references within). Our present proposal demands, however, an extension to the multi-impurity scenario. For this we rely on the intuition that the impurities can be ``fused'', to advance some jargon, in an effective boundary perturbation. In this case, we introduce a \emph{multi-impurity fusion ansatz} allowing writing elegant expressions for correlation functions and impurity entropies. Instead of pursuing a formal demonstration of this ansatz, we then take a more pedestrian approach: we exploit toy-models of increasing complexity which can, nevertheless, be solved exactly to provide direct examples of our hypothesized results. We believe this approach is more pragmatic, allowing a circumvention of abstract concepts like boundary states. Our examples consist of multi-impurity versions of scalar scattering, single-channel Kondo -- both Abelian and somewhat trivial scenarios -- and culminate on an extension of the Emery-Kivelson solution~\cite{Emery_Kivelson} of the anisotropic 2CK problem (at the Toulouse point) to the multi-impurity case. 

As a final remark, we call attention to the known relationship between CFT and fusion category theories, used to describe anyons and topological quantum computation. The fractionalized particles in NCK differ from anyons in topological phases in that they are bound to the impurities, similarly as Majoranas are bound to edges in Kitaev chains. In the above criterion for ``universal utility'' of anyons, they are confined to defects and therefore are not prone to thermally activated generation of pairs. On the other hand, these particles cannot be braided. Since we can extract the quantum information stored in the impurities via correlation functions, we offer some final remarks regarding an implementation of measurement-only topological quantum computation~\cite{Bonderson_topo_comp} to substitute braiding operations on this system.

The paper is organized as follows: in Sec.~\ref{sec:model} we discuss the chiral multi-impurity multi-channel Kondo model as well as a preliminary venue to implement it. We explore subtleties of this proposal and comment on possible ways around them. In Sec.~\ref{sec:fusion_ansatz} we describe the multi-fusion ansatz which, we argue, solves the chiral Kondo problem of the previous section. We comment on the relationship between this ansatz and fusion category theories, discuss how it affects the spectrum and correlation functions of chiral multi-impurity problems. Finally we offer examples of application and realization of the ansatz. We conclude in Sec.~\ref{sec:conclusion}, with remarks on implementations of measurement-only computation as well as branching directions of our work.

\section{The Chiral multi-channel multi-impurity Kondo model \label{sec:model}}

The chiral multi-channel and multi-impurity (MCMI) Kondo model is described by a Hamiltonian with two pieces, $H_{\rm{MCMI}}=H_0+H_K$. The first part, $H_0$ describes a set of $N$ spinful chiral electrons,
\begin{equation}
    H_{0}=-\frac{v_F}{2\pi}\int dx\psi^{\dagger}i\partial_{x}\psi, \label{eq:Kins}
\end{equation}
where $\psi$ is a spinor of components $\psi_{i,\sigma}$, with $i=1,...,N$ for channel and $\sigma=\uparrow,\downarrow=\pm$ for spin (our notation will vary according to necessity). We consider the electrons to be right movers, for concreteness. The second part of the Hamiltonian, $H_K$, contains interactions with a sequence of $M$ magnetic impurities 

\begin{align}
    H_{K}&=g\int dx\frac{1}{2}\left(\psi^{\dagger}\boldsymbol{\sigma}\psi\right)\left(x\right)\cdot \boldsymbol{\mathcal{S}}\left(x\right) \nonumber \\
    \boldsymbol{\mathcal{S}}\left(x\right)&=\sum_{l=1}^{M}\delta\left(x-x_{l}\right)\left(\begin{array}{c}S_{l,x}\\S_{l,y}\\\Delta S_{l,z}\end{array}\right) \label{eq:Konds}
\end{align}
where $x_l<x_{l+1}$, $g$ is a coupling constant, and $\Delta$ is an anisotropy parameter. The impurities have spin-$1/2$. Spin anisotropy is  irrelevant to the low-energy NFL behavior of our interest.~\cite{Anderson_Kondo,Affleck_anisotropy} Nevertheless, the spin anisotropy will be important for the exact solution of the 2CK problem explored below.~\cite{Emery_Kivelson} In contrast, perfectly symmetric interactions between the channels and the impurities are crucial to achieve the NFL low-energy behavior.~\cite{Affleck_anisotropy} In practice  NFL behavior occurs at a quantum critical point tuned by symmetry breaking perturbations, and, has a finite stability to the latter at finite energy scales.~\cite{pustilnik2004quantum,sela2011exact,mitchell2012universal}

For the single impurity problem, the Kondo effect is characterized by the Kondo temperature $T_K=\Lambda e^{-1/g}$, where $\Lambda$ is an ultraviolet scale, and an associated long length scale $\xi_K=\hbar v_F/(k_B T_K)$. For long distances and low energies, $\xi_K$ controls the localization length of the electrons that screen the impurity.~\cite{Affleck_Kondo_cloud} We therefore will be interested in a length scale hierarchy
\begin{equation}
    \left|x_{l+1}-x_{l}\right|\gg \xi_K. \label{eq:Lhierar}
\end{equation} 

The interest in  this ``dilute'' scenario of Eq.~\eqref{eq:Lhierar} is to work in a situation in which the impurities do not correlate with each other. This will be a pivotal point in the calculation of the spectrum of the MCMI problem. As discussed below, in this limit, the spectrum calculation follows a sequential application of the fusion ansatz, frequently and successfully employed in the determination of the low-energy fixed point spectrum of the NCK problem.~\cite{Affleck_Kondo_review} Diluteness, however, is not enough to guarantee the independence of the local moments. In principle, long-range RKKY interactions between the impurities should fully lift the ground state degeneracy. Non-trivially, we find that it is a feature, due to the chiral nature of the system, that RKKY interactions vanish identically here. Intuitively, this can be seen by noting that RKKY interactions between two local magnetic moments require them to mutually exchange information via conduction electrons. If the electronic bath is chiral, the information can only flow in a single direction and no actual exchange coupling can be established.~\footnote{Mathematically this can be verified perturbatively. The RKKY coupling is most easily computed for Kondo couplings smaller than the temperature of the system, where divergences of free imaginary-time integrals return just a temperature factor. For the chiral problem, the rest of the calculation returns a coefficient controlled by integrals of the form $\int d\tau 1/(\tau+i x)^2$, where $x$ is the distance between the impurities. Since both poles are on the same side of the complex plane, the integral vanishes. This argument can be extended to any order in perturbation theory, so that the coupling exactly vanishes. } The final consequence of this decoupled scenario is that in this limit anyon-like objects are well-defined at each impurity, as in the single-impurity case. These bound states are isolated from each other and, jointly, they define a bona fide anyon Hilbert space in which quantum information may be stored non-locally. The manipulation of this quantum information may possibly be done in a non-invasive manner through non-demolition measurements.~\cite{Bonderson_topo_comp}

Before continuing to our solution of the problem and its implications, we propose an attempt at realizing the MCMI Hamiltonian. This proposal contains the main ingredients in $H_{\rm{MCMI}}$. We discuss, however, some of the difficulties we are aware of, leaving a more refined modelling for future considerations.

\begin{figure}[tb]
	\centering
	\includegraphics[width=1.0\columnwidth]{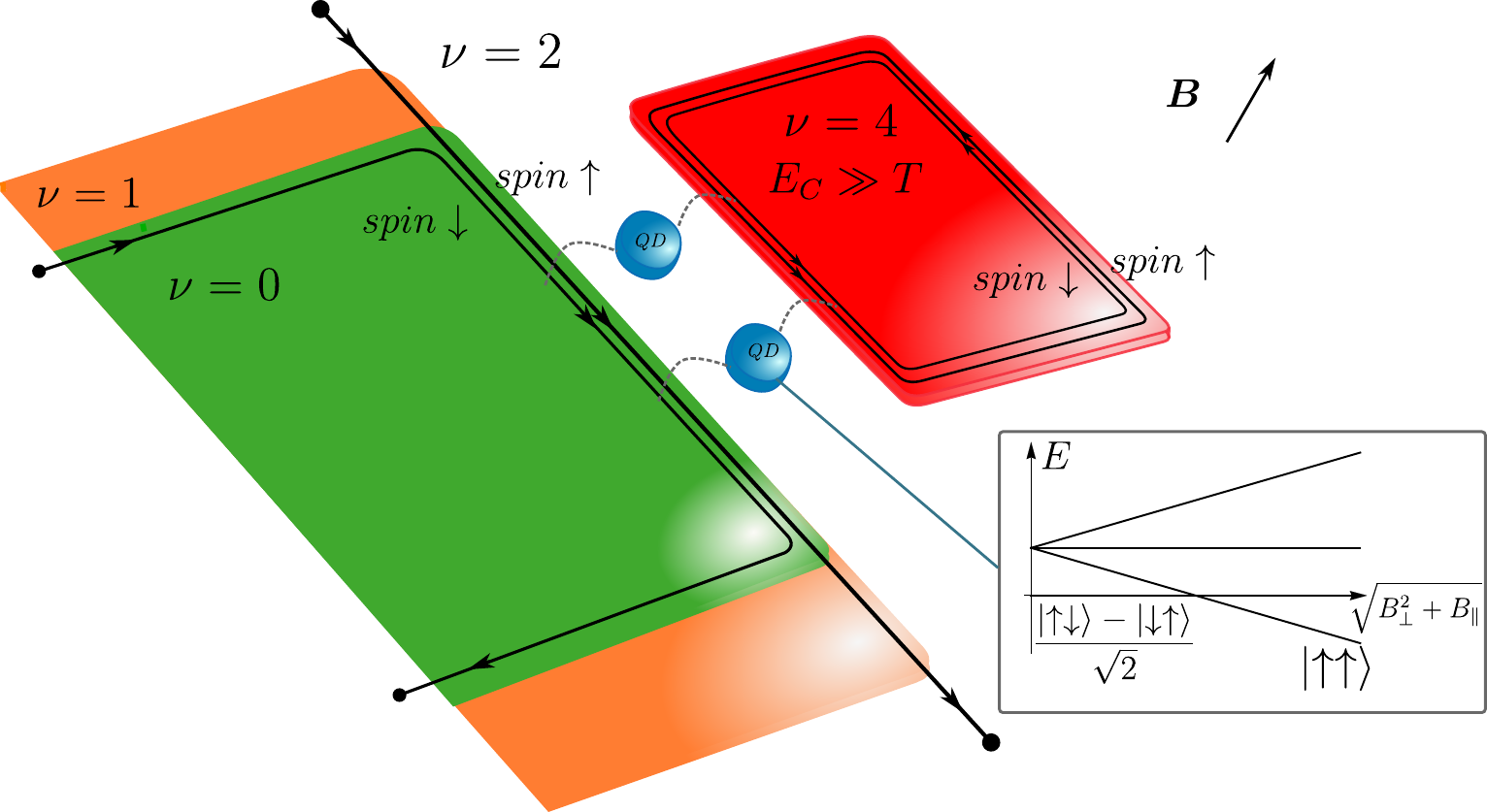}
	\caption{Chiral, 2-impurity, 2-channel Kondo device schematics. The phenomenology of interest can be engineered by pairs of doubly degenerate integer quantum Hall edges with quantum dots (QDs). The QDs' degenerate two-particle energy levels in the presence of a magnetic field act as  effective magnetic impurities while distinct channels correspond to distinct quantum Hall islands (red, green). Doping may be used to split the quantum Hall channels for cross correlation measurements.}
	\label{fig:Le_device}
\end{figure}

The system we have in mind is depicted in Fig.~\ref{fig:Le_device}, for the case of a 2-channel, 2-impurity, chiral Kondo system. We consider a 2D electron gas in the presence of a strong perpendicular magnetic field $B_\perp$. The role of the chiral channels is played by edge modes of integer quantum Hall (IQH) phases. Three main IQH domains are defined by doping: regions in the phases of magnetic filling fractions $\nu=0,\,2,$ and $4$ (green, white, and red, respectively). Every boundary between domains experiences a jump of $\Delta n_{\rm{TKNN}}=2$ in the TKNN invariant -- Chern number -- evincing the existence of two conducting channels at each boundary.~\cite{TKNN} These conducting channels are completely filled Landau levels of spin-polarized electrons, one with spin up and the other with spin down. So a given boundary gives us a pair of spinful chiral electrons. The device displays two such boundaries, the edges between the green and white regions and white and red regions. Overall, we thus get two channels of chiral of spinful electrons, fixing $N=2$ in the Hamiltonian.

Inside the white $\nu=2$ region we lay a pair of quantum dots (QD, blue). For more impurities, one may simply imagine more dots. Proposals for the NCK effect in QDs in the presence of a magnetic field require using dot states of even charge~\cite{Pustilnik_evenQ_Kondo,Kikoin_oreg_2CK}; we focus on two-particle states. Coulomb repulsion splits the two-particle states in the dot between a singlet and triplet. The applied magnetic field, however, Zeeman-splits the triplet state. With an in-plane magnetic field component $B_\parallel$, this splitting can be tuned until the $\left|\uparrow\uparrow\right\rangle $ triplet state becomes degenerate with the singlet. The result is a two-level system which can be hybridized by spin flips; proximitizing, the QDs act as magnetic impurities for the spinful IQH edges. The 2CK phase can only be achieved if the green and red electronic reservoirs are independent from each other, individually -- but simultaneously and symmetrically -- attempting to screen the impurities. To preclude cooperative impurity screening by the two channels, electron transfer processes between the reservoirs must be suppressed.~\cite{oreg2003two,Potok_GG_2CK} We imagine doing so by considering the red reservoir to have a large capacitive energy $E_{\rm{C}}$ exceeding the temperature $T$, precluding charge build-up that could have been absorbed from the green reservoir. As a final detail, the orange $\nu=1$ doped region has the purpose of splitting the two spinful channels, so that their currents could be measured in separate and cross correlations can be studied.

While this device captures the main details of the MCMI Hamiltonian, we point out some possible implementation subtleties. Since the whole device operation relies on the IQH effect, the spin degeneracy in the spinful channels is not guaranteed; in fact it is not expected. 
The effects of this asymmetry are discussed in detail in Appendix~\ref{sec:dev_toy}. Essentially, our analysis shows that the absence of spin symmetry is reflected in a spin-dependent tunneling of electrons from the edge modes to the QD. This results in induced perturbations, the most relevant of them being an effective local magnetic field removing the degeneracy of the impurity. Fortunately, this term may be tuned via $B_\parallel$ to zero under the condition of degeneracy of the two renormalized QD levels, showing some robustness of our proposal. Unfortunately, the magnetic field over the conduction electrons remains which cannot be simultaneously tuned away, but this term has a lesser impact in the Kondo effect. The possibility of persistence of 2CK physics in a similar model has been further discussed in  Ref.~\onlinecite{Kikoin_2CK_evenQ}. Turning shortly our attention to the green and red channels to the QDs, we note that channel asymmetry would be lethal to the NFL regime, but can be trivially accounted for by the system geometry. 

Given the points above, a careful engineering of this device may be necessary, but many solutions to the problems are possible. The issue of spin asymmetry, for example, may be also dealt with by exploiting another quantum number such as the valley degeneracy of graphene quantum Hall edge modes. 
Another possibility would be to rely on the edge modes of Chern insulators. Finally, designs more closely based on charge Kondo approaches usually display more robust signatures. This happens as these systems operate in a tunneling resonance regime which make the Kondo coupling to be characterized by the hopping amplitude, rather than exchange coupling of the spin Kondo applications, hence enhancing the Kondo energy scale. 

A relatively minor caveat of this geometry would be the generalization of this device to more channels. For example, designing a model to realize 3CK would be highly desirable in the future, given the prospects of obtaining Fibonacci anyons at the impurities of the 3CK phase. We will proceed with our solution and analysis of the model, leaving a finer device modelling for future ventures.

\section{Solution of the MCMI problem \label{sec:fusion_ansatz}}

The NCK fixed point (FP) physics is captured by CFT methods. This is not surprising at the decoupled FP (no Kondo interactions), which simply contains chiral free fermions. At the strongly interacting FP, however, the CFT phenomenology called for the introduction of a ``fusion ansatz'', which has been extensively verified for single and two impurities by comparison with numerical RG.~\cite{Affleck_2IK_NRG} Its consequences also agree with the spectrum found via Bethe ansatz.~\cite{Andrei_NCK_Bethe,Furuya_NCK_Bethe} Let us review this discussion, as it sets the context and motivates the introduction of the multi-fusion ansatz below.

The CFT approach to the NCK problem involves a sequence of conformal embedding identifications. Considering spin and channels, one has a total of $2N$ fermions. First, by non-Abelian bosonization, one identifies the non-interacting FP with two decoupled CFTs: $U(1)\times SU(2N)_1$. This is nothing but a generalized version of spin-charge separation. The $U(1)$ part corresponds to a charge degree of freedom by means of a boson field, while $SU(2N)_1$ is a Wess-Zumino-Witten (WZW) CFT. One then proceeds with a conformal embedding based on the level-rank duality~\cite{BYB}
\begin{equation}
    SU\left(2N\right)_{1}=SU\left(N\right)_{2}\times SU\left(2\right)_{N}.
\end{equation}
In this decomposition, the $SU(2)_N$ sector contains the degrees of freedom carrying the spin quantum numbers. This is where the fusion ansatz takes place. It ascribes the strongly interacting FP to a primary-field fusion of the $SU(2)_N$ sector with a $j=1/2$ primary field, corresponding to the impurity. The $U(1)$ and $SU(N)_2$ sectors remain pristine.

This process is more elegantly captured in the framework of Cardy's BCFT.~\cite{CARDY_1,CARDY_2,Cardy_review} Recalling the features of BCFT demands a short detour which will pay off later in this paper. In this methodology, local perturbations that preserve conformal symmetry are introduced at the two edges of a finite-sized (non-chiral length $L/2$) 1D system. The effects of these perturbations are abstractly associated with a pair of boundary states $\left|A\right\rangle ,\:\left|B\right\rangle$. Phenomenologically, the effect of boundary conditions can be seen at statistical quantities. We focus here on two particular ones:

(i)\emph{The partition function} - by conformal invariance, partition functions are decomposed in terms of characters associated with (conformal) towers of primary states/fields $\Phi_a$. If $\chi_a(q)$ is the character of a chiral theory on a torus, with periodic space of length $L$ and imaginary time with circumference given by the inverse temperature, the partition function for a 
system with boundary conditions $A$ and $B$ reads
\begin{equation}
    Z_{AB}=\sum_{a}n_{AB}^{a}\chi_{a}\left(q\right), \label{gluing}
\end{equation}
where $q=e^{ -\frac{ 2 \pi v_F }{LT}}$. The multiplicities $n_{AB}^{a}$ are ``gluing conditions''. Physically, conformal boundary conditions  affect the spectrum only through $n_{AB}^{a}$. The characters $\chi_a(q)$ are determined from the torus geometry, ignoring boundaries. The latter only enter in the statistical mechanics through $n_{AB}^{a}$, determining which conformal tower contribute to $Z_{AB}$.

(ii)\emph{The two-point correlation functions} - let us assume the chiral unfolded picture with boundary perturbations $B$ at the origin, the right-most extremity of a system, and send $L \to \infty$. For complex coordinates $z=\tau-ix$, describing \emph{right movers},
the correlations between primary fields evaluated at points ${\rm{Im}}(z_1)<0$ and ${\rm{Im}}(z_2)>0$ around the impurity are changed from the standard CFT result only by a multiplicative constant~\cite{Cardy_Lewellen}
\begin{equation}
    \left\langle \Phi_{a}\left(z_{1}\right)\Phi_{a}\left(z_{2}\right)\right\rangle =\frac{1}{\left(z_{1}-z_{2}\right)^{2h_{a}}}\frac{\left\langle \left.a\right|B\right\rangle }{\left\langle \left.0\right|B\right\rangle }, \label{2pcor}
\end{equation}
where $h_a$ is the conformal dimension of the primary $\Phi_a$. The constant $\left\langle \left.a\right|B\right\rangle /\left\langle \left.0\right|B\right\rangle $ is a ratio of amplitudes between boundary states and the highest weight primary states $\left|0\right\rangle ,\left|a\right\rangle $ of the vacuum, $\Phi_0$,  and $\Phi_a$ conformal towers.

By a clever manipulation using conformal invariance and the so-called $S$-modular-transformation which reverses the generators of a torus, it is possible to relate gluing conditions and the amplitudes featuring in the two-point correlation functions through the so-called Cardy's equations~\cite{Cardy_review}
\begin{equation}
    \left\langle \left.A\right|b\right\rangle \left\langle \left.b\right|B\right\rangle =\sum_a n_{AB}^{a}S_{a}^{b}, \label{eq:CardyEqs}
\end{equation}
where $S_{a}^{b}$ is a matrix representation of the modular $S-$transformation.~\cite{BYB}
The issue with Cardy's equations is that, a priori, one has no information about either of its sides. This is where the fusion ansatz comes in. It first associates a primary-field $\Phi_c$ to a boundary condition $B\equiv B(c)$. Then it determines that the distinction between a system with both free boundaries $\left|F\right\rangle ,\:\left|F\right\rangle$ and one with a single non-trivial boundary $\left|F\right\rangle ,\:\left|B\right\rangle $ is given by a change in the gluing condition as
\begin{equation}
    n_{FB}^{a}=\sum_{b}N_{bc}^{a}n_{FF}^{b}. \label{eq:fusion}
\end{equation}
Here $N_{bc}^{a}$ are the coefficients of the regular operator product expansion, or fusion, algebra of CFT primary fields:
\begin{equation}
    \Phi_{a}\times\Phi_{b}=\sum_{c}N_{ab}^{c}\Phi_{c}. \label{eq:CFTfusion}
\end{equation}


Manipulating Cardy's equation together with Verlinde's formula,~\cite{VERLINDE} one then obtains a non-trivial result: the ratio of amplitudes that fix the correlation functions are determined by the modular $S$-matrix
\begin{equation}
    \frac{\left\langle \left.a\right|B\right\rangle }{\left\langle \left.0\right|B\right\rangle }=\frac{S_{c}^{a}/S_{0}^{a}}{S_{c}^{0}/S_{0}^{0}}.
\end{equation}
Further extending this analysis allows the determination of all correlation functions of the NCK problem by fusing the $SU(2)_N$ sector with a $c\equiv j=1/2$ primary field due to a Kondo boundary state $\left|K\right\rangle \equiv \left|B\right\rangle_{c=\frac{1}{2}}$ [\emph{c.f.} Fig.~\ref{fig:Beff}(a)]. The point we approach below is how to implement the BCFT framework to the MCMI problem.


\subsection{Multi-fusion hypothesis and relation to non-abelian anyons}

Formulas such as Cardy's equations, Eq.~\eqref{eq:CardyEqs}, and the 2-point correlation functions, Eq.~\eqref{2pcor}, contain the full information necessary to characterize the physics of the Kondo FP. The derivation of these formulas, however, relies on the existence of a center of inversion in the problem. That is usually the origin, where the boundary perturbation is applied and around which one may fold or unfold space ({\emph{c.f.}} Fig. \ref{fig:Kondos}). This becomes an issue at the MCMI problem, with its spaced $M$ impurities.

A solution to the MCMI problem would require a novel implementation of BCFT. Instead of considering a system with a boundary, one needs to ask what are the effects of several conformal-symmetry-preserving localized perturbations. 
A much simpler way out arises if one relies, instead on a physical motivation. As long as the impurities are far apart and do not interact with each other, we may expect that their effect on the chiral degrees of freedom is 
independent. Patches are separated by the impurities and the modes are sequentially affected as they interact locally with each impurity. This procedure is captured by a ``multi-fusion'' ansatz, as described by a natural generalization of Eq.~\eqref{eq:fusion} reading
\begin{equation}
    n_{FM}^{a}=\sum_{b}n_{FM-1}^{b}N_{bc}^{a}.
\end{equation}
Here, $n_{FM}^{a}$ represents the gluing conditions for a chiral system subjected to a free boundary condition on its left, and $M$ localized equal perturbations as it moves to its right. This gluing condition is simply obtained from the same gluing condition for $M-1$ perturbing impurities by use of the fusion-rule coefficients. Defining $M=0\equiv F$, we may iterate this equation, leading to
\begin{equation}
   n_{FM}^{a}=\sum_{b_{1},...,b_{M}}n_{FF}^{b_{1}}\left(N_{b_{1}c}^{b_{2}}N_{b_{2}c}^{b_{3}}...N_{b_{M-1}c}^{b_{M}}N_{b_{M}c}^{a}\right).
\end{equation}
This equation has a useful consequence. To understand it, we make connection with anyon physics or, more precisely, with fusion category theory. Here, bringing anyons $a$ and $b$ together leads to a set of fusion outcomes $c$ according to the fusion rules
\begin{equation}
    a\times b=\sum_{c}N_{ab}^{c}c,
\end{equation}
which index a set of quantum states $\left|a,b;c\right\rangle $. We can relate a 2D anyon fusion theory to a corresponding CFT related to the modes at the 2D edges (this correspondence is not one-to-one).~\cite{KITAEV_honey,BaisSlingerland,dubail2012edge} In this case, the anyon fusion coefficients are the same as the primary-field fusion rules in Eq.~\eqref{eq:CFTfusion}. As is well-known, we can relate the fusion coefficients to the dimension of the Hilbert space of anyon fusion. For example, fusing $M$ anyons of type $c$, we may write
\begin{align}
    \underbrace{c\times c\times....\times c}_{M}&=\sum_{c_{1},...,c_{M-2},c_{\rm{eff}}}N_{cc}^{c_{1}}N_{c_{1}c}^{c_{2}}...N_{c_{M-2}c}^{c_{\rm{eff}}}c_{\rm{eff} } \nonumber\\&=\sum_{ c_{\rm{eff}}}\text{dim}\left[V_{c...c}^{c_{\rm{eff}}}\right]c_{\rm{eff}},
\end{align}
where $\text{dim}\left[V_{c...c}^{c_{\rm{eff}}}\right]$ is the dimension of the Hilbert space obtained by $(M-1)$ fusions of anyons $c$ into the anyon $c_{\rm{eff}}$. This dimension is dominated by the anyon $c$ quantum dimension $d_c$, the highest eigenvalue of the matrix of entries $N^b_{ca}\equiv (N_c)_{ab}$.

With this background, one may use the associativity of the anyon fusion algebra to obtain
\begin{equation}
    \left(\left(\left(\left(b_{1}\times c\right)\times c\right)\times...\right)\times c\right)=\sum_{b_{2},...,b_{M},a}N_{b_{1}c}^{b_{2}}N_{b_{2}c}^{b_{3}}...N_{b_{M}c}^{a}a,
\end{equation}
from one side, and 
\begin{align}
&\left(b_{1}\times\left(\left(\left(c\times c\right)\times...\right)\times c\right)\right)\nonumber \\&=b_{1}\times\left(\sum_{c_{1},...,c_{M-2},c_{\rm{eff}}}N_{cc}^{c_{1}}N_{c_{1}c}^{c_{2}}...N_{c_{M-2}c}^{c_{\rm{eff}}}c_{\rm{eff}}\right)\nonumber \\&=\sum_{c_{1},...,c_{M-2},c_{\rm{eff}},a}N_{cc}^{c_{1}}N_{c_{1}c}^{c_{2}}...N_{c_{M-2}c}^{c_{\rm{eff}}}N_{b_{1}c_{\rm{eff}}}^{a}a \label{eq:fusion_match}
\end{align}
from another. Returning to the multi-fusion ansatz it can now be rewritten as
\begin{align}
    n_{FM}^{a}&\equiv\sum_{c_{\rm{eff}}}\text{dim}\left[V_{c...c}^{c_{\rm{eff}}}\right]n_{FB_{\rm{eff}}}^{a},
\end{align}
where we used the regular fusion ansatz to write
\begin{equation}
    n_{FB_{\rm{eff}}}^{a}=\sum_{b}n_{FF}^{b}N_{bc_{\rm{eff}}}^{a}.
\end{equation}
\begin{figure}[tb]
	\centering
	\includegraphics[width=1.0\columnwidth]{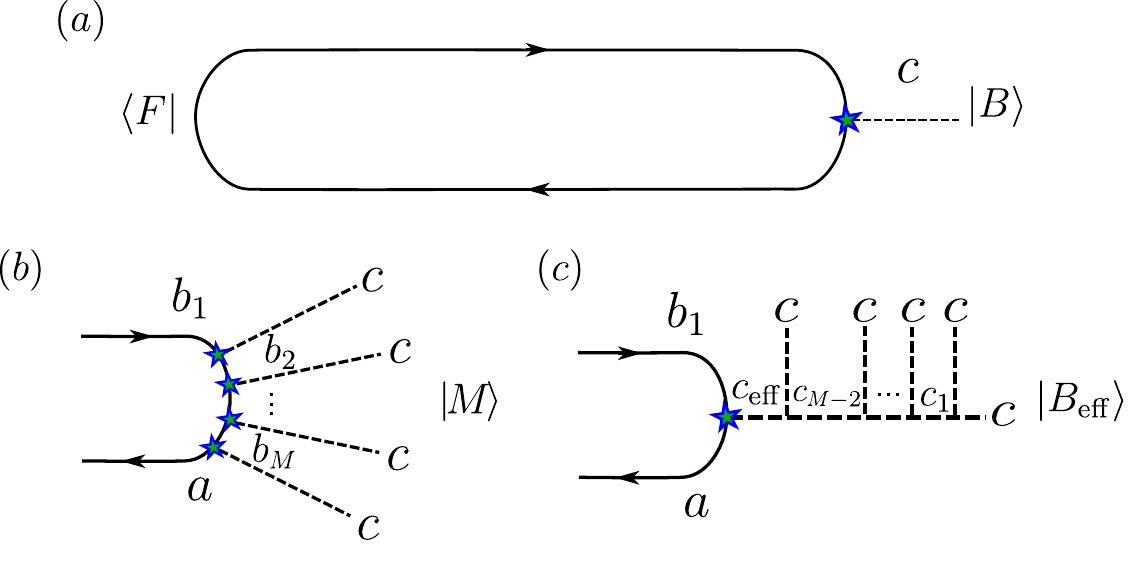}
	\caption{Fusion ansatz schematics. (a)  $M$ chiral fermions are bounded in a closed system of circumference $L$. A trivial $F$ boundary condition is applied on one side and a non-trivial boundary condition $B$ (in the Kondo case, due to a spin-1/2 magnetic impurity) on the other side. The boundary condition $B$ is equivalent to fusing an anyon $c$ to the conduction fermions. 
	(b) Particles in a chiral sector $b_1$ cross a set of $M$ primary $c$ perturbations. At each interaction the particles change by fusion with $c$, emerging at the end as particles of type $a$. (c) We reinterpret the sequential fusion of $c$ anyons, exchanging it for a single anyon $c_{\rm{eff}}$ that results from a set of equal-anyonic fusions in parallel.} 
	\label{fig:Beff}
\end{figure}
The two equivalent processes of fusion in Eq.~\eqref{eq:fusion_match} are depicted graphically in Figs.~\ref{fig:Beff}(b) and (c). We exchanged the set of all impurities by a single effective impurity corresponding to an anyon $c_{\rm{eff}}$. To it one associates a primary field and a boundary state $\left|B_{\rm{eff}}\right\rangle $ which, as before, depends on the primary field $\Phi_{c_{\rm{eff}}}$, the effective boundary perturbation. The cost is an extra statistical multiplicative factor scaling with the quantum dimension of the anyons $\sim d_{c}^{M-1}$. 

Now, for a single effective impurity, we may deploy the regular toolbox of BCFT. For instance, the partition function of the system, Eq.~\eqref{gluing}, becomes
\begin{align}
    Z_{FM}&=\sum_{a}n_{FM}^{a}\chi_{a}\left(q\right) \nonumber \\&=\sum_{a}\sum_{c_{\rm{eff}}}\text{dim}\left[V_{c...c}^{c_{\rm{eff}}}\right]n_{FB_{\rm{eff}}}^{a}\chi_{a}\left(q\right) \nonumber \\&=\sum_{c_{\rm{eff}}}\text{dim}\left[V_{c...c}^{c_{\rm{eff}}}\right]Z_{FB_{\rm{eff}}}.
\end{align}
For a single impurity, the sum and ``dim'' drop out, returning the usual result. For multiple impurities, physical processes are weighted by all the possibilities of outcome from fusion of all impurities. This leads, for example, to a multiplicative factor of $M$ in the impurity entropy, in accordance with the extensive nature of entropy.

The exchange between the sequential fusions to a global effective fusion from Figs.~\ref{fig:Beff}(b) to (c) has an impact in the correlation functions. The effective fusion channel implies the existence of a convenient basis of conserved quantum numbers for the computation of correlation functions. The sequential fusion of the chiral modes leads to scattering into distinct primary-field superselection sectors. The effective fusion picture in Fig.~\ref{fig:Beff}(c) simply organizes these independent sectors. For a given one, the two-point correlation function reads
\begin{align}
\left\langle \Phi_{a}\left(z_{1}\right)\Phi_{a}\left(z_{2}\right)\right\rangle_{c_{\mathrm{eff}}} &=\frac{1}{\left(z_{1}-z_{2}\right)^{2h_{a}}}\frac{\left\langle \left.a\right|B_{\rm{eff}}\right\rangle }{\left\langle \left.0\right|B_{\rm{eff}}\right\rangle } \nonumber \\&=\frac{1}{\left(z_{1}-z_{2}\right)^{2h_{a}}}\frac{S_{c_{\rm{eff}}}^{a}/S_{0}^{a}}{S_{c_{\rm{eff}}}^{0}/S_{0}^{0}}.    \label{eq:effcorr}
\end{align}
In other words, since all fusion channels contribute to the partition function, one expects correlations to have contributions from all $c_\mathrm{eff}$ sectors too. Eq.~\eqref{eq:effcorr} are building blocks of the total fermion correlator. We will interpret these building blocks as probabilistic outcomes of the correlation measurement, depending on the fusion channel associated with $c_{\rm{eff}}$. In this case, two-point correlation-function in a given fusion sector may access topological anyon information, as entries of the modular $S-$matrices. 

In the next section we  provide evidence that the multi-fusion ansatz is indeed satisfied in cases of interest, most notably the MCMI problem.

\subsection{Examples}

To support the validity of the multi-fusion ansatz, we consider a few simple examples.

\subsubsection{Scalar Scattering}

We start in the simplest scenario of interest: scalar scattering of chiral spinless fermions. While this problem does not relate directly to the Kondo problem, it allows for complete solution and visualization of the effects of the multi-fusion ansatz [in a $U(1)$ scenario], as well as nuances about regimes of its validity. 

Consider a system with finite size $x \in [-L/2,L/2] $ and spinless chiral electrons in the presence of a set of local potential scatterers
\begin{align}
    H&=\int_{-L/2}^{L/2}dx\psi^{\dagger}\left[-i\partial_{x}+2\pi\sum_{l=1}^{M}V^{\left(l\right)}\delta\left(x-x_{l}\right)\right]\psi.
\end{align}
We assume antiperiodic boundary conditions on the fermions, $\psi\left(L/2\right)=-\psi\left(-L/2\right)$, for concreteness, and order the scatterer positions as $x_l<x_{l+1}$ with the sole constraint of being overall confined to a small subregion of the whole system, {\emph{i.e.}} $|x_M-x_1|\ll L$. Otherwise the scatterers are spread apart in any arbitrary way.

To compute the eigenmodes of this problem and, particularly, their Green's function, we perform a gauge transformation
\begin{equation}
    \psi=e^{i2\pi\sum_{l=1}^{M}V^{\left(l\right)}\theta\left(x-x_{l}\right)}\tilde{\psi}.
\end{equation}
Now, $\tilde{\psi}$ obeys a free Hamiltonian, but has boundary  condition
\begin{equation}
    \tilde{\psi}\left(L/2\right)=-e^{-i2\pi\sum_{l=1}^{M}V^{\left(l\right)}}\tilde{\psi}\left(-L/2\right).
\end{equation}
The free fermion $\tilde{\psi}$ admits a momentum space expansion but, due to the new boundary conditions, momentum quantization returns
\begin{equation}
    k_{m}=\frac{2\pi}{L}\left[\left(m+\frac{1}{2}\right)-\sum_{l=1}^{M}V^{\left(l\right)}\right],\quad m\in\mathbb{Z}.
\end{equation}
Overall, the eigenmode expansion of the original fermions then reads
\begin{align}
\psi&=\frac{e^{i2\pi\sum_{l=1}^{M}V^{\left(l\right)}\theta\left(x-x_{l}\right)}}{\sqrt{L}}\sum_{m}e^{ik_{m}x}a_{m} \nonumber \\&=\frac{e^{i2\pi\sum_{l=1}^{M}V^{\left(l\right)}\left[\theta\left(x-x_{l}\right)-\frac{x}{L}\right]}}{\sqrt{L}}\sum_{m}e^{i\frac{2\pi}{L}\left(m+\frac{1}{2}\right)x}a_{m},
\end{align}
for whose energy state creation and annihilation operators $a^\dagger_m,\,a_m$, we associate eigenenergies $E_m=k_m$.

We remark on involved length scales. If we introduce a length scale $\xi_{K}$ over which the scattering through each impurity occurs, then we would like to assume $x_{l}-x_{l+1}>\xi_{K}$. This way we can define a region of size $\left(M-1\right)\xi_{K}$ and call it an effective boundary. For $\xi_{K}\ll L$ and a finite number of impurities ($M\ll L/\xi_{K}$), we can imagine that this effective boundary barely affects the size of the system, {\emph{i.e.}}, the effective system would have a size $L'=L-\left(M-1\right)\xi_{K}\approx L$. In that case, evaluating the field operator at a position $x<x_{1}$ or $x>x_{M}$ but with $x\ll L$ would lead to operators accumulating a phase difference of roughly $\alpha_{tot}=2\pi\sum_{l=1}^{M}V^{\left(l\right)}$. In fact, if the sum of the distance between the impurities is small with respect to the system size $L$, one can see that the fermion field simply ``eats'' successively the impurities, one at a time from each site. This is in agreement with an Abelian sequential fusion of each impurity.

As we see, the effects of the scalar scatterings on the operators is twofold. First, the eigenfunctions phases jump discontinuously at each impurity. Second, their momenta are shifted. The whole scale discussion above is much simplified if one takes the thermodynamic limit $L\to\infty$. In this case
\begin{equation}
    \psi\left(x\right)=e^{i2\pi\sum_{l=1}^{M}V^{\left(l\right)}\theta\left(x-x_{l}\right)}\int_{-\infty}^{\infty}\frac{dk}{2\pi}e^{ikx}a\left(k\right), \label{eq:contPSI}
\end{equation}
where $a\left(k\right)=\sqrt{L}a_{m}$ and the energies read $E\left(k\right)=k$. The momentum shifts drop out from our expressions. With this expression, the 2-point Green's function can be computed explicitly, reading
\begin{equation}
    G\left(\tau,x,\tau',x'\right)=\frac{e^{i2\pi\sum_{l=1}^{M}V^{\left(l\right)}\left[\theta\left(x-x_{l}\right)-\theta\left(x'-x_{l}\right)\right]}}{z-z'}.
\end{equation}
Let us stress that this expression is valid on the plane, {\emph{i.e.}} in the $L\to\infty,\,T\to0$ limit. In Appendix~\ref{sec:sca_green} we provide a more in-depth discussion on subtleties related to calculations on the torus and the thermodynamic limit. For the moment, we just stress  that this is the exact result expected by the multi-fusion ansatz. The boundary state amplitudes for this U(1) theory are controlled by simple phase factors which accumulate sequentially as impurities are crossed. Fixing the positions of $x$ and $x'$ relative to the scatterers shows that they behave as conformally invariant boundary conditions (that is, they affect the CFT correlator only by a multiplicative factor). If $x<x_1$ and $x'>x_M$, we absorb the total phase $\alpha_{tot}$ of the effective boundary. Notice that, nevertheless, measuring correlations between points in between the scatterer positions allows each of the distinct phases $V^{(l)}$ to be determined separately.

\subsubsection{NCK fixed points via CFT \label{sec:FP_CFT}}

Now we start exploring the problem in the actual Kondo context. We start with Eqs.~\eqref{eq:Kins} and~\eqref{eq:Konds}, for $N$ channels and several $M$ impurities. Our goal will be to show that for a special choice of coupling constants, and in the thermodynamic limit, the arguments that lead to the usual fusion ansatz also lead to the multi-fusion ansatz.~

We set $v_F=1$ and anisotropy parameter $\Delta=1$ and, again, use complex coordinates for the chiral fermions: $\psi=\psi\left(x,\tau\right)=\psi\left(z=\tau-ix\right)$. Defining $U(1)$, $SU(2)$ and $SU(N)$ current operators
\begin{align}
J\left(z\right) =:\psi^{\dagger}\psi:,~~~
\vec{J}\left(z\right)  =:\psi^{\dagger}\frac{\boldsymbol{\sigma}}{2}\psi:,~~~J^{A}\left(z\right)=:\psi^{\dagger}T^{A}\psi:,
\end{align}
with $SU(N)$ generators $T^A$ acting on the flavor indices, we rewrite the Hamiltonian density in Sugawara  form as~\cite{AFFLECK_Kondo_1990,BYB}
\begin{align}
\mathcal{H}_{0}&=\frac{1}{8\pi N}J^{2}+\frac{1}{2\pi\left(N+2\right)}\left|\vec{J}\right|^{2}+\frac{1}{2\pi\left(N+2\right)}J^{A}J^{A}
\end{align}
and
\begin{equation}
\mathcal{H}_{K}=\sum_{l=1}^{M}g_{l}\vec{J}\cdot\vec{S}_{l}\delta\left(x-x_{l}\right).
\end{equation}
Since interactions only couple to the spin sector, we focus on it.
At a finite size $L$, we Fourier decompose,
\begin{align}
\vec{J}_{n}=\frac{1}{2\pi}\int_{-L/2}^{L/2}dxe^{i\frac{2\pi}{L}nx}\vec{J}\left(x\right), \,\,\,\vec{J}\left(x\right)=\frac{2\pi}{L}\sum_{n}e^{-i\frac{2\pi}{L}nx}\vec{J}_{n}. 
\end{align} 
The Fourier components satisfy the $SU(2)_N$ Kac-Moody algebra
\begin{equation}
\left[J_{n}^{a},J_{m}^{b}\right]=N\frac{n}{2}\delta^{ab}\delta_{n+m,0}  +i\epsilon^{abc}J_{n+m}^{c}.
\end{equation}
Defining
\begin{equation}
\vec{\mathcal{J}}_{n}=\vec{J}_{n}+\frac{N+2}{2}\sum_{l=1}^{M}e^{i\frac{2\pi}{L}nx_{l}}g_{l}\vec{S}_{l},
\end{equation}
the spin part of the Hamiltonian becomes
\begin{equation}
    H_{s}=\frac{2\pi}{(N+2)L} \sum_{n}\vec{\mathcal{J}}_{-n}\cdot\vec{\mathcal{J}}_{n}+{\rm{constant}}.
\end{equation}
What is the algebra obeyed by $\vec{\mathcal{J}}_{n}$?  By computing the commutators,
\begin{align}
 \left[\mathcal{J}_{n}^{a},\mathcal{J}_{m}^{b}\right]&=N\frac{n}{2}\delta^{ab}\delta_{n+m,0}  \\&+i\epsilon^{abc}\left[J_{n+m}^{c}+\left(\frac{N+2}{2}\right)^{2}\sum_{l=1}^{M}g_{l}^{2}e^{i\frac{\pi}{L}\left(n+m\right)x_{l}}S_{l}^{c}\right]. \nonumber
\end{align}
By imposing
\begin{equation}
\frac{N+2}{2}g_{l}=1\Rightarrow g_{l}=\frac{2}{N+2}\:\forall l,
\end{equation}
a similar constraint as found in the single impurity case,~\cite{AFFLECK_Kondo_1990}  we recover an $SU(2)_N$ Kac-Moody algebra. 

For this fine-tuned coupling, the spin sector of the theory is described by an SU(2) symmetric WZW CFT, just as in the decoupled problem.~\cite{Affleck_Kondo_review} The new spin operators,
\begin{equation}
\vec{\mathcal{J}}_{n}=\vec{J}_{n}+\sum_{l=1}^{M}e^{i\frac{\pi}{L}nx_{l}}\vec{S}_{l},
\end{equation}
show that for this coupling the impurity spins are simply absorbed by the conduction degrees of freedom. The total spin is nothing but the $n=0$ Fourier component of the spin current,
\begin{equation}
\vec{\mathcal{J}}_{0}=\vec{J}_{0}+\sum_{l=1}^{M}\vec{S}_{l}, \label{eq:spindens}
\end{equation}
and demonstrates simple angular momentum addition. As in the single impurity problem, $g_l=2/(N+2)$ is representative of the strong-coupling fixed point.  

The physical interpretation of these results is straightforward. In the absence of the impurity, every state is characterized by $J_{0}^{2}=j\left(j+1\right)$,
or simply spin $j$. At the strong-coupling fixed point, this spin changes according to~\eqref{eq:spindens}. We have to sum the angular momentum with all the impurities. Since the fine-tuned -- or equivalently, the strongly coupled -- fixed point has a spectrum still controlled by an $SU(2)_{N}$ theory, the standard truncation of WZW CFTs takes place.

Let us look at an example, the $N=1$ single channel Kondo problem. As usual, the conformal towers $(Q,j)$ for charge and spin sectors are constrained by gluing conditions enforcing, at the decoupled fixed point,
\begin{equation}
\left({\rm{even}},\mathbb{Z}\right)\oplus\left({\rm{odd}},1/2+\mathbb{Z}\right).
\end{equation}
Meanwhile, the strong-coupling gluing conditions change with the number of impurities. We observe

\begin{align}
\left({\rm{even}},\mathbb{Z}\right)\oplus\left({\rm{odd}},1/2+\mathbb{Z}\right)\ {\rm{for}}\ M\:{\rm{even}}\\
\left({\rm{even}},1/2+\mathbb{Z}\right)\oplus\left({\rm{odd}},\mathbb{Z}\right)\ {\rm{for}}\ M\:{\rm{odd}}.
\end{align}
Naturally, this suggests that the strongly coupled fixed point is characterized, far from the impurities, by free electrons up to phase shifts. These are given by the number of impurities times $\pi/2$.

These results are in agreement with the multi-fusion ansatz from the effective boundary CFT picture, from which we can extract the phase shifts explicitly. Indexing primary fields by the total spin $j$, the fusion rules of $SU(2)_1$ read
\begin{gather}
0\times0=0,\;~~\frac{1}{2}\times\frac{1}{2}=0 , ~~~ \;0\times\frac{1}{2}=\frac{1}{2}.
\end{gather}
For an odd number of impurities, this natural truncation of the $SU(2)_1$ Kac-Moody algebra enforces  $c_{\rm{eff}}=1/2$. For an even number of impurities, $c_{\rm{eff}}=0$. The modular $S-$matrix of $SU(2)_1$ reads
\begin{equation}
    S=\left(\begin{array}{cc}
\frac{1}{\sqrt{2}} & \frac{1}{\sqrt{2}}\\
\frac{1}{\sqrt{2}} & -\frac{1}{\sqrt{2}}
\end{array}\right).
\end{equation}
Since fermions are primary fields of spin $j=1/2$ and conformal dimension $h_{1/2}=1/2$, Eq.~\eqref{eq:effcorr} recovers
\begin{equation}
    \left\langle \psi_{\sigma}\left(z_{1}\right)\psi_{\sigma}^{\dagger}\left(z_{2}\right)\right\rangle =\frac{1}{z_{1}-z_{2}}\begin{cases}
\frac{S_{1/2}^{1/2}/S_{0}^{1/2}}{S_{1/2}^{0}/S_{0}^{0}}=-1 & M\, \mathrm{odd}\\
\frac{S_{0}^{1/2}/S_{0}^{1/2}}{S_{0}^{0}/S_{0}^{0}}=1 & M\, \mathrm{even}
\end{cases},
\end{equation}
which displays the $\pi/2$ phase shifts for odd number of impurities.   

The analysis we just made can be repeated for higher number of channels, by fixing the proper fusion rules and $S-$matrices. Since the two-channel and three-channel scenarios are also independently noteworthy, we will turn our attention to these cases next. 

\subsubsection{Exact results in the 2CK case}
Using the Emery-Kivelson construction,~\cite{Emery_Kivelson} we can also make some explicit progress in the 2CK case. For rigor and details, we follow more closely the approach of von Delft \emph{et al}.~\cite{vondelft_2CK} Our goal here is twofold. First, we will show that the multi-impurity problem is consistent with the existence of localized Majorana fermions at each impurity site. Second, we will show that we can extract a parity factor from the electron-electron correlation function that is in agreement with the multi-fusion ansatz.

\emph{Decoupled Majoranas} -- The reasoning behind the Emery-Kivelson approach is as follows. The natural physical degrees of freedom of the 2CK problem are given by the number of electrons of given spin and channel
\begin{equation}
    \hat{\tilde{\mathcal{N}}}_{i\sigma}=\int\frac{dx}{2\pi} \psi_{i\sigma}^{\dagger}\psi_{i\sigma},~~~(i=1,2,~~\sigma = \uparrow, \downarrow). 
\end{equation}
The corresponding states $\left|\vec{\mathcal{\tilde{N}}}\right\rangle =\left|\mathcal{N}_{1\uparrow},\mathcal{N}_{1\downarrow},\mathcal{N}_{2\uparrow},\mathcal{N}_{2\downarrow}\right\rangle$, however, are inconvenient due to strong fluctuations of particle numbers in each flavor. A much more convenient basis is constructed by an orthogonal transformation
\begin{align}
\vec{\mathcal{N}}&\equiv\left(\begin{array}{c}
\hat{\mathcal{N}}_{c}\\
\hat{\mathcal{N}}_{s}\\
\hat{\mathcal{N}}_{cf}\\
\hat{\mathcal{N}}_{sf}
\end{array}\right) \label{eq:Orot} =\frac{1}{2}\left(\begin{array}{cccc}
1 & 1 & 1 & 1\\
1 & -1 & 1 & -1\\
1 & 1 & -1 & -1\\
1 & -1 & -1 & 1
\end{array}\right)\left(\begin{array}{c}
\hat{\tilde{\mathcal{N}}}_{1\uparrow}\\
\hat{\tilde{\mathcal{N}}}_{1\downarrow}\\
\hat{\tilde{\mathcal{N}}}_{2\uparrow}\\
\hat{\tilde{\mathcal{N}}}_{2\downarrow}
\end{array}\right) 
\end{align}
into charge, spin, charge-flavour, and spin-flavor sectors, which we write simply as $\hat{\mathcal{N}}_{\eta}\equiv\mathcal{O}_{\eta,i\sigma}\hat{\tilde{\mathcal{N}}}_{i\sigma}$.

In this new basis the charge and charge-flavor sectors decouple, since they are exactly conserved. Once including the impurity spins, conservation of total angular momentum also provides a strong constraint. The electronic spin $s$ can flip by unity only, and only locally due to interactions with the impurities. Albeit large fluctuations can happen globally, locally they are mild; large spin fluctuations can only happen in high-order processes. In contrast with the other degrees of freedom, however, the spin-flavor $sf$ sector can fluctuate more wildly. This fluctuation lies at the roots of the NFL behavior of the 2CK strongly coupled fixed point.~\cite{vondelft_2CK} Next we see that the Emergy-Kivelson solution generalizes nicely to the $M$-impurity scenario.

To be able to rotate the basis as in Eq.~\eqref{eq:Orot}, we bosonize the fermions,
\begin{align}
    \psi_{i\sigma}\left(x\right)&=\frac{\kappa_{i\sigma}}{\sqrt{a_{0}}}e^{-i\tilde{\Phi}_{i\sigma}\left(x\right)}\\\tilde{\Phi}_{i\sigma}\left(x\right)&\equiv\left(\hat{\tilde{\mathcal{N}}}_{i\sigma}-\frac{1}{2}\right)\frac{2\pi x}{L}+\tilde{\phi}_{i\sigma}\left(x\right),
\end{align}
where $a_0$ is a UV cutoff and antiperiodic boundary conditions were chosen on a finite-size system. Our convention will be that zero-mode pieces are singled-out from the lower case boson variables $\tilde{\phi}_{i\sigma}$, but included in the capitalized ones, as $\tilde{\Phi}_{i\sigma}$. The Klein-factors obey $\left[\kappa_{i\sigma},\hat{\mathcal{N}}_{j\sigma'}\right]=\delta_{ij}\delta_{\sigma \sigma'}\kappa_{i\sigma}$, $\left\{ \kappa_{i\sigma},\kappa_{j \sigma'}\right\} =2\delta_{ij}\delta_{\sigma \sigma'}$, and $\kappa_{i\sigma}^{\dagger}\kappa_{i\sigma}=1$. The bosons also obey equal-time commutation relations
\begin{equation}
    \left[\Phi_{\eta}\left(x\right),\Phi_{\eta'}\left(y\right)\right]=i\pi\delta_{\eta\eta'}\text{sgn}\left(x-y\right). \label{eq:boson_commut}
\end{equation} 

In the boson language, it is straightforward to change basis as $\Phi_{\eta}\equiv\mathcal{O}_{\eta,i\sigma}\tilde{\Phi}_{i\sigma}$. Fixing Klein factors is slightly trickier but not different from as in Ref.~\onlinecite{vondelft_2CK}. For example, it is possible to convince oneself that
\begin{equation}
    \kappa_{sf}^{\dagger}\kappa_{s}^{\dagger}\equiv\kappa_{1\uparrow}^{\dagger}\kappa_{1\downarrow},
\end{equation}
as well as 
\begin{align}
    \kappa_{sf}\kappa_{s}^{\dagger}&=\kappa_{2\uparrow}^{\dagger}\kappa_{2\downarrow}\\\kappa_{sf}^{\dagger}\kappa_{cf}^{\dagger}&=\kappa_{1\uparrow}^{\dagger}\kappa_{2\uparrow},
\end{align}
where $\kappa_\eta$ are the new Klein factors obeying $\left\{ \kappa_{\eta},\kappa_{\eta'}^{\dagger}\right\} =2\delta_{\eta\eta'}$, $\left[\kappa_{\eta},\hat{\mathcal{N}}_{\eta'}\right]=\delta_{\eta\eta'}\kappa_{\eta}$,  and $\left[\kappa_{\eta},\Phi_{\eta'}\right]=0$.  The definition of new Klein-factors is possible as the physical Hilbert space where the Hamiltonian acts is restricted to contain only states that satisfy the gluing conditions and are closed under the pairwise action of Klein-factors. We remark, crucially, that the number of Klein-factors arising from the conduction electrons does not change with the number of impurities. This is of course natural, but has implications, as usually the localized impurity Majoranas stem from properties of these Klein factors. After the considerations above, the 2-channel multi-impurity Hamiltonian becomes
\begin{align}
   H_{0}&=\frac{\Delta_{L}}{2}\sum_{\eta}\hat{\mathcal{N}}_{\eta}^{2}+\sum_{\eta}\frac{1}{2}\int\frac{dx}{2\pi}\left(\partial_{x}\phi_{\eta}\right)^{2}\\H_{K,z}&=g_{z}\sum_{l=1}^M\left[\partial_{x}\phi_{s}\left(x_{l}\right)+\Delta_{L}\hat{\mathcal{N}}_{s}\right]S_{l,z}\\H_{K,\perp}&=\frac{g_{\perp}}{2a_{0}}\sum_{l=1}^{M}\left(\kappa_{sf}^{\dagger}e^{i\Phi_{sf}\left(x_{l}\right)}+\kappa_{sf}e^{-i\Phi_{sf}\left(x_{l}\right)}\right) \nonumber \\&\times\left(S_{l,-}\kappa_{s}^{\dagger}e^{i\Phi_{s}\left(x_{l}\right)}-S_{l,+}\kappa_{s}e^{-i\Phi_{s}\left(x_{l}\right)}\right),
\end{align}
where $g_{\perp} \equiv g $, $g_z =\Delta g$, $\Delta_{L}=2\pi/L$. We explicitly separated the $z$ couplings from the ``perpendicular'' couplings of the Kondo interaction in Eq.~\eqref{eq:Konds} and, with the hindsight of the CFT-based calculation from the previous section, we took the interactions to be equal in strength at all impurities.

We succeeded in separating the charge and flavor-charge degrees of freedom from the spin ones. To move on, the spin and spin-flavor can be decoupled by a unitary transformation at the Toulouse point $g_z=1$. This unitary transformation reads, in the present scenario
$\mathcal{U}=U_1...U_M$, where 
\begin{equation}
    U_{l}=e^{ig _{z}S_{l,z}\Phi_{s}\left(x_{l}\right)}.
\end{equation}
In applying $\mathcal{U}$ to the Hamiltonian, one has to take into account an ordering of the impurity positions, as well as the anomalous chiral boson commutation relations~\eqref{eq:boson_commut}. A careful consideration of this point and refermionization in terms of
\begin{equation}
    \psi_{\eta}\left(x\right)=\frac{\kappa_{\eta}}{\sqrt{a_{0}}}e^{-i\left(\hat{\mathcal{N}}_{\eta}-\frac{1}{2}\right)\frac{2\pi x}{L}}e^{-i\phi_{\eta}\left(x\right)}
\end{equation}leads to the Hamiltonian $H'=\mathcal{U} H \mathcal{U}^{-1}$
\begin{align}
H_{0}^{'}&=\sum_{\eta}\int\frac{dx}{2\pi}\psi_{\eta}^{\dagger}i\partial_{x} \psi_{\eta}\\H_{K,\perp}^{'}&=\frac{g_{\perp}}{2}\sum_{l=1}^{M}\left[e^{i\frac{\pi x_{l}}{L}}\psi_{sf}^{\dagger}\left(x_{l}\right)+e^{-i\frac{\pi x_{l}}{L}}\psi_{sf}\left(x_{l}\right)\right]\left(d_{l}-d_{l}^{\dagger}\right). \nonumber
\end{align}
The complex fermions $d_l$ here read
\begin{equation}
    d_{l}=\kappa_{s}^{\dagger}S_{l,-}e^{i\pi\left(\sum_{m=l+1}^{M}S_{m,z}\right)}.
\end{equation}
The Jordan-Wigner-like strings appear here after the consideration of boson commutations relations mentioned above. They are an encouraging surprise. Both the Klein-factor and the spin strings are absolutely crucial to guarantee anti-commutation between fermions defined at distinct impurities, as well as canonical anti-commutation with their conjugates. Also we notice again the finite-size oscillatory factors which would lead to commensurability effects. Taking the $L\to \infty$ limit and introducing a Majorana basis
\begin{align}
    \psi_{\eta}=\frac{\chi_{\eta}^{1}+i\chi_{\eta}^{2}}{\sqrt{2}},~~~~~d_{l}=\frac{a_{l}+ib_{l}}{\sqrt{2}},
\end{align}
we obtain the final form of the problem:
\begin{align}
    H_{0}^{'}&=\frac{i}{2}\sum_{\eta,a=1,2}\int\frac{dx}{2\pi}\chi_{\eta}^{a}\left(x\right)\partial_{x}\chi_{\eta}^{a}\left(x\right)\\H_{K,\perp}^{'}&=ig_{\perp}\sum_{l=1}^M\chi_{sf}^{1}\left(x_{l}\right)b_{l}. \label{eq:H2CK}
\end{align}
As usual, the 2CK problem at the Toulouse point is described by a set of four independent fermionic fields, only one of which remains coupled to the impurity. This coupling is now of a single-particle nature, and is written most naturally in the Majorana basis. The problem is exactly solvable. Above all, we note the existence of $M$ decoupled Majorana fermions $a_l$, one at each impurity. The standard picture of a single impurity is again reproduced in series, suggesting the validity of the multi-fusion ansatz.

Solving for the eigenmodes of this problem displays other signatures of the multi-fusion ansatz. For example, it can be shown that the $b_l$ Majorana modes are absorbed by the conduction electrons. Details of this calculation and analysis can be found in Appendix~\ref{sec:2CK_eig_app}.

\emph{Correlation function and parity factors} -- Upon scattering at a spin-1/2 impurity, electrons in the 2CK regime are completely transformed into collective degrees of freedom associated with the spinor representation of an underlying $SO(8)$ symmetry, evident due to the 4 complex fermions, or 8 Majoranas in the free problem.~\cite{MALDACENA_Ludwig_SO8} The simplest consequence of this is that electron-electron correlation functions vanish identically for an odd number of impurities. We will, therefore, consider an even number of impurities, and ask ourselves if we can extract information from two-point correlation functions that is in agreement with the multi-fusion expectation. 

Consider the correlation function
\begin{align}
    &\left\langle \psi_{i\sigma}\left(z\right)\psi_{j\sigma'}^{\dagger}\left(z'\right)\right\rangle \nonumber \\&=\frac{1}{a_{0}}\left\langle \kappa_{i\sigma}\kappa_{j\sigma'}^{\dagger}e^{-i\tilde{\Phi}_{i\sigma}\left(z\right)}e^{i\tilde{\Phi}_{j\sigma'}\left(z'\right)}\right\rangle \nonumber \\&=\frac{1}{a_{0}}\left\langle \kappa_{i\sigma}\kappa_{j\sigma'}^{\dagger}e^{-i\mathcal{O}_{i\sigma,\eta}\Phi_{\eta}\left(z\right)}e^{i\mathcal{O}_{j\sigma',\eta'}\Phi_{\eta'}\left(z'\right)}\right\rangle.
\end{align}
To extract any information related to the fusion channel of the effectively decoupled impurity Majoranas $a_l$, we have to perform the same set of manipulations as in the transformed Hamiltonian discussed above. This is a delicate procedure. We must perform the $\mathcal{U}$ unitary transformation at the Toulouse $g_z=1$ point, but ought to be careful with commutation relations being defined at equal times. Only the spin sector is affected by this procedure, so we focus on $\mathcal{O}_{i\sigma,s}=\sigma/2$. Assuming $x<x_{1}$ and $x'>x_{M}$, we have
\begin{align}
    \mathcal{U}e^{-i \frac{\sigma}{2}\Phi_{s}\left(x\right)}\mathcal{U}^{-1}&=e^{-i \frac{\sigma}{2}\Phi_{s}\left(x\right)-i \frac{\pi \sigma}{2}S_{z,\rm{imp}}} \nonumber \\\mathcal{U}e^{i \frac{\sigma'}{2} \Phi_{s}(x')}\mathcal{U}^{-1}&=e^{i \frac{\sigma'}{2}\Phi_{s}(x')-i\frac{\pi \sigma'}{2} S_{z,\rm{imp}}},
\end{align}
where $S_{z,\rm{imp}}=\sum_{l=1}^M S_{l,z}$. Since the impurity spin is not conserved by itself, however, one has to be careful before plugging these results back in the correlation function. A way around this is to remember that the  $a_l$ fermions  are fully decoupled from the Hamiltonian~\eqref{eq:H2CK}. So, we write
\begin{align}
    e^{-i \frac{\pi \sigma}{2} S_{z,\rm{tot}}} =&\left(-1\right)^{\frac{\sigma}{2} \sum_{l=1}^M\left(d_{l}^{\dagger}d_{l}-\frac{1}{2}\right)} \nonumber \\=&\left(-1\right)^{\frac{\sigma M}{4} }\prod_{l=1}^M\left(2ia_{l}b_{l}\right)^{\frac{\sigma}{2}}.
\end{align}
We define the parity of the state of $M$ fused Majoranas of type $a$
\begin{align}
  \mathcal{P}_a=(2i)^{M/2} \prod_{l=1}^M a_l,
\end{align}
 {\emph{i.e.}} their fusion channel, and similarly, $\mathcal{P}_b=(2i)^{M/2} \prod_{l=1}^M b_l$. This gives $\prod_{l=1}^M\left(2a_{l}b_{l}\right) = \mathcal{P}_a \mathcal{P}_b$.
Note finally that the Klein-factors enforce the spin and channel labels to be equal in the correlation function, giving two identical factors of the form $e^{-i \frac{\pi \sigma}{2} S_{z,\rm{tot}}}$, that $\left(\mathcal{P}_{a,b}\right)^{2\mathcal{O}_{i\sigma,s}}=\mathcal{P}_{a,b}$, and that $\mathcal{P}_a$ can commute through the Hamiltonian. Also, the charge, flavour and spin channels are decoupled and their correlations can be obtained from the operator product expansions of the corresponding vertex operators. We obtain the final expression
\begin{widetext}

\begin{equation}
\left\langle \psi_{j\sigma}\left(z\right)\psi_{j\sigma}^{\dagger}\left(z'\right)\right\rangle \propto   \frac{\mathcal{P}_{a}}{\left(z-z'\right)^{3/4}}\left\langle e^{-i\mathcal{O}_{j\sigma,sf}\Phi_{sf}\left(z\right)} \left[P_{b}\left(\tau\right)\right]^{\sigma/2}e^{i\mathcal{O}_{j\sigma,sf}\Phi_{sf}\left(z'\right)}\left[P_{b}\left(\tau'\right)\right]^{\sigma/2}\right\rangle, 
\end{equation} 
where we used $\mathcal{O}_{j\sigma,c}^{2}+\mathcal{O}_{j\sigma,f}^{2}+\mathcal{O}_{j\sigma,s}^{2}=3/4$. We achieved the goal of extracting exactly the parity operator $\mathcal{P}_a$ from the correlation function. We cannot, unfortunately, make more rigorous progress than this. The extra power of $1/4$ hides in the remaining correlation function where the $b$-parity operator and remaining bosonic exponentials are not expected to commute. If one accepts some educated guesswork, however, we may resort to adapting the -- rigorous, albeit ignoring Klein factors and decoupled Majorana sectors -- arguments of Maldacena and Ludwig here.~\cite{MALDACENA_Ludwig_SO8} In their work, which goes beyond the Toulouse limit, the effect of an impurity scattering in the strongly coupled 2CK regime is connected to a change in boundary conditions whereby the spin-flavor boson flips sign. If we generalize this to our many-impurity scenario, we expect
\begin{equation}
    \Phi_{sf}\left(z\right)|_{{\rm{Im}}\left(z\right)=x_{l}-\epsilon}=\left(-1\right)^{l}\Phi_{sf}\left(z\right)|_{{\rm{Im}}\left(z\right)=x_{l}+\epsilon}.
\end{equation}
Together with our results, it is not far-fetched to presume that one may absorb the $b$-Majorana parities into the $\Phi_{sf}$ bosons by a change in their boundary conditions. At the end one would simply have the CFT result for the vertex operator correlator
\begin{align}
    \left\langle e^{-i\mathcal{O}_{j\sigma,sf}\Phi_{sf}\left(z\right)} \left[P_{b}\left(\tau\right)\right]^{\sigma/2}e^{i\mathcal{O}_{j\sigma,sf}\Phi_{sf}\left(z'\right)}\left[P_{b}\left(\tau'\right)\right]^{\sigma/2}\right\rangle \nonumber \to \frac{1}{\left(z-z'\right)^{1/4}}.
\end{align}
\end{widetext}
This way, the electronic 2-point function for an even number of impurities would reduce to
\begin{equation}
    \left\langle \psi_{j\sigma}\left(z\right)\psi_{j\sigma}^{\dagger}\left(z'\right)\right\rangle_a =\frac{\mathcal{P}_{a}}{z-z'},
\end{equation}
and this would be in agreement with the multi-fusion ansatz. The fusion rules for $SU(2)_2$ read
\begin{gather}
    0\times0=0,\;1\times1=0,\;\frac{1}{2}\times\frac{1}{2}=0+1 \nonumber\\1\times\frac{1}{2}=\frac{1}{2},\;0\times\frac{1}{2}=\frac{1}{2},\;0\times1=1.
\end{gather}
The corresponding $S-$matrix is 
\begin{equation}
    S=\left(\begin{array}{ccc}
\frac{1}{2} & \frac{1}{\sqrt{2}} & \frac{1}{2}\\
\frac{1}{\sqrt{2}} & 0 & -\frac{1}{\sqrt{2}}\\
\frac{1}{2} & -\frac{1}{\sqrt{2}} & \frac{1}{2}
\end{array}\right).
\end{equation}
For an odd number of impurities, $c_{\rm{eff}}=1/2$ and, as $S^{1/2}_{1/2}=0$, the correlation function should vanish. For an even number of impurities, however, $c_{\rm{eff}}=0$ or $c_{\rm{eff}}=1$, corresponding to $\mathcal{P}_a=1$ or $\mathcal{P}_a=-1$, respectively. 
This can be written as
\begin{align}
    \left\langle \psi_{j\sigma}\left(z_{1}\right)\psi_{j\sigma}^{\dagger}\left(z_{2}\right)\right\rangle =\frac{1}{z_{1}-z_{2}}\begin{cases}
\frac{S_{1/2}^{1/2}/S_{0}^{1/2}}{S_{1/2}^{0}/S_{0}^{0}}=0 & M\,\rm{odd}\\
\frac{S_{0}^{1/2}/S_{0}^{1/2}}{S_{1/2}^{0}/S_{0}^{0}}=1 & M\,\rm{even}\\
\frac{S_{1}^{1/2}/S_{0}^{1/2}}{S_{1}^{0}/S_{0}^{0}}=-1 & M\,\rm{even}
\end{cases}.
\end{align}
The coefficient $M$-even cases is nothing but $\mathcal{P}_a$, as mentioned. The factor of $-1$ in the $c_{{\rm{eff}}}=1$ channel suggests a $\pi/2$ phase-shift associated with a two-channel screening of a spin-1 effective particle. The two values of $\mathcal{P}_a$, and consequently of the correlation function suggest a double ground state degeneracy, due to the free Majoranas. 
In sum, the 2CK problem has been worked out rigorously here, to the point where $\mathcal{P}_a$ is extracted from the correlation function. This is the main signature expected from the multi-fusion ansatz. 
The most noteworthy lesson here comes from contrasting with the 1CK case. In the 2CK problem we see that even for a fixed number of impurities, two distinct solutions appear for the correlation function. These solutions, corresponding to two anyon fusion channels, come in the form of simple phase shifts, but that stems from the fact that the two outcomes of the fusion $1/2\times 1/2=0+1$, namely $0$ and $1$, are still Abelian anyons.

\subsubsection{3CK and Fibonnaci anyons}


We finish our discussion by pointing the non-trivial consequences if we push this picture to the 3CK scenario. Now, the low energy theory is expected to be controlled by an $SU(2)_3$ WZW CFT. We emphasize again the existence of a relationship between this CFT and Fibonacci anyons, first alluded in the introduction. While we cannot solve the 3CK problem exactly, we may rely on our ansatz, supported by the strong-coupling fixed point results of Subsection~\ref{sec:FP_CFT}. The modular $S-$matrix reads
\begin{equation}
    S=\left(\begin{array}{cccc}
\sqrt{1-\frac{1}{\sqrt{5}}} & \sqrt{1+\frac{1}{\sqrt{5}}} & \sqrt{1+\frac{1}{\sqrt{5}}} & \sqrt{1-\frac{1}{\sqrt{5}}}\\
\sqrt{1+\frac{1}{\sqrt{5}}} & \sqrt{1-\frac{1}{\sqrt{5}}} & -\sqrt{1-\frac{1}{\sqrt{5}}} & -\sqrt{1+\frac{1}{\sqrt{5}}}\\
\sqrt{1+\frac{1}{\sqrt{5}}} & -\sqrt{1-\frac{1}{\sqrt{5}}} & -\sqrt{1-\frac{1}{\sqrt{5}}} & \sqrt{1+\frac{1}{\sqrt{5}}}\\
\sqrt{1-\frac{1}{\sqrt{5}}} & -\sqrt{1+\frac{1}{\sqrt{5}}} & \sqrt{1+\frac{1}{\sqrt{5}}} & -\sqrt{1-\frac{1}{\sqrt{5}}}
\end{array}\right).
\end{equation}
The fusion rules for $SU(2)_3$ are more cumbersome. For simplicity, we focus on the $1/2\times 1/2=0+1$ fusion rule, which is unchanged in comparison with the 2CK case above, and $M=2$ impurities. The prefactors of the correlation functions, however, are now completely different. If the impurities fuse to $c_{\rm{eff}}=0$, we just have a trivial boundary condition and a unity prefactor. If, however, $c_{\rm{eff}}=1$, the correlation function prefactor now reads $-(\sqrt{5}-1)/(\sqrt{5}+1)$. Not only they are different in sign, but also in modulus. There is no full conversion of electronic degrees of freedom into other collective modes. Scattering by an odd number of impurities does not need to vanish in this case.

\section{Summary and Outlook \label{sec:conclusion}}

We proposed a chiral multi-channel and multi-impurity Kondo model as a venue for realizing quasiparticle fractionalization and anyon-related physics. Contrasting with platforms based on standard topologically ordered phases, our system is gapless. Our anyonic modes are separated from the conduction electron fluid due to  inherent frustration present in the multi-channel Kondo effect, leading to residual strongly correlated and spatially localized fractionalized degrees of freedom.

\begin{figure}[tb]
	\centering
	\includegraphics[width=1.0\columnwidth]{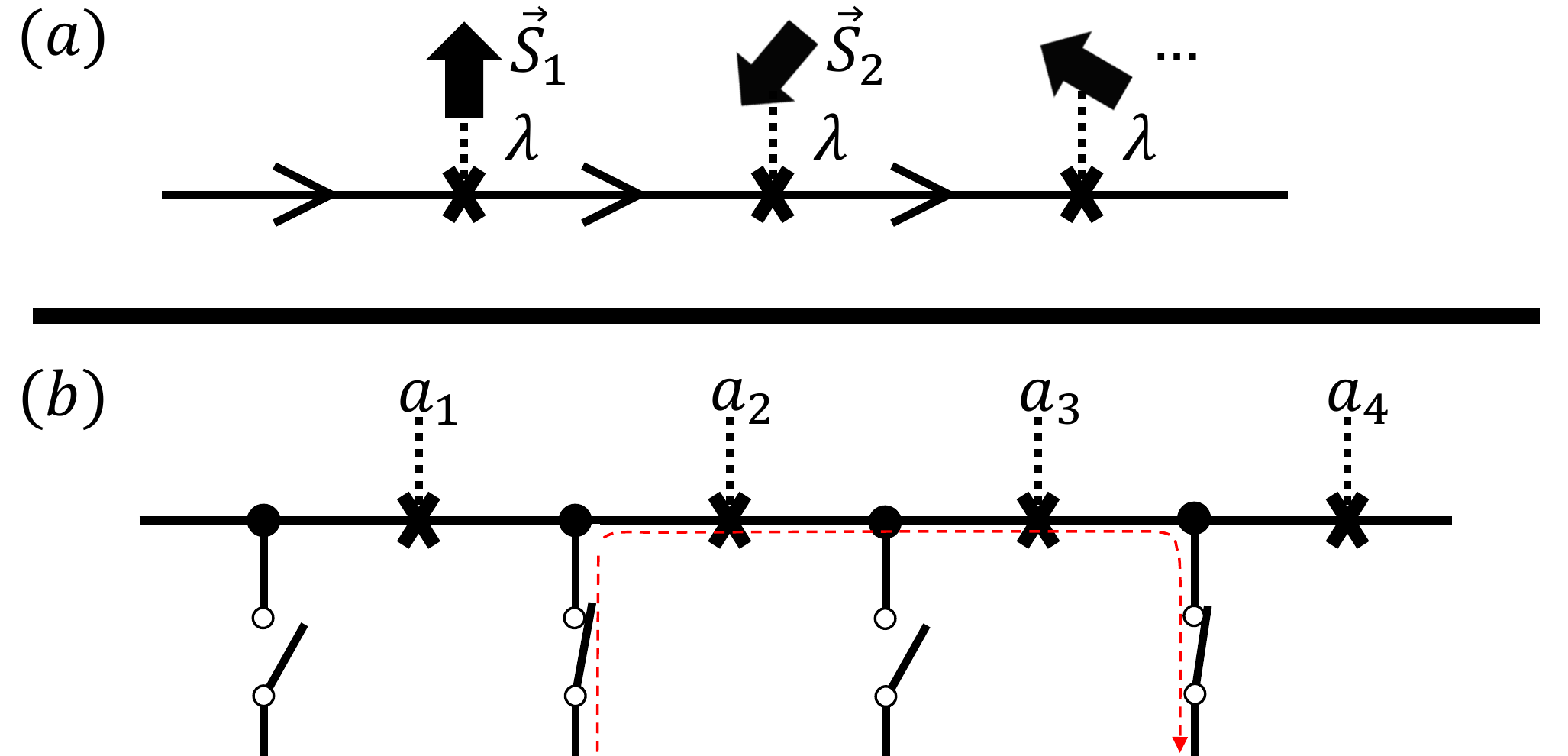}
	\caption{Simplified picture of the chiral Kondo Model (a) and its overscreened anyon low-energy picture (b). Switches between impurities allow the extraction of electrons at any particular position. This allows the measurement of correlation functions extracting information about a fusion channel of a desired sequence of anyons at their corresponding impurities.}
	\label{fig:Switch}
\end{figure}

The implementation described here is a promising platform for manipulation of topological quantum information. Through a multi-fusion adaptation of the boundary conformal field theory fusion ansatz, we have shown that impurity-anyon fusion channels leave signatures in two-point correlation functions. While the static impurity fractional particles can't be braided, the signatures left on correlation functions provide the very requirements for the implementation of measurement-only topological quantum computation.~\cite{Bonderson_topo_comp} In this methodology, anyon braiding is substituted by measurements of anyon charges, which in this context mean simply the result of a fusion channel. No direct interaction between impurities is necessary for the manipulation of the information. Computations are determined in a probabilistic fashion, determined by the result of repeated measurements of fusion outcomes. These depend only on the interaction between the conduction electrons and each impurity, and can be selected through switches, as illustrated in Fig.~\ref{fig:Switch}.

If the  building blocks of the correlation function (associated with fixed fusion channel $c_\mathrm{eff}$)  can indeed be separately measured, a simplified protocol for quantum information manipulation can be devised. One starts with a system of $M$ impurities, equipped with the possibility to inject an electron before the $l$-th impurity and extract it after the $m$-th ($m>l$) impurity  [red dashed line in Fig.~\ref{fig:Switch}(b)]. This allows to measure the fusion outcome of all anyons in the segment from $l$ to $m$. Our 1D setup is limited to measuring the fusion of anyons along connected segments. Still, this  allows interesting operations such as state teleportation. While success in achieving teleportation by such method is only probabilistic, the number of measurements returning failed processes is exponentially suppressed.~\cite{Bonderson_topo_comp} To realize effective braiding of the anyons using measurements of fusion outcomes requires to go beyond our purely 1D system,~\cite{Bonderson_topo_comp,vijay2016teleportation} effectively allowing us to measure the fusion outcome of nonconsecutive anyons on the line. For this purpose we envision adding external coherent connectors on our chiral channel, extracting an electron from one point and reinjecting it from another. 

Chiral multi-channel Kondo systems then provide an exciting venue for studying quasiparticle fractionalization. They satisfy the standard primitives for the realization of topological quantum computation:~\cite{Bonderson_topo_comp} they display (i) non-Abelian particles whose (ii) fusion channel may be determined by correlation function measurements. Furthermore, the anyons are bound to defects, which offers a beneficial extra protection of the information from thermal proliferation of anyon pairs.~\cite{UFQP} In comparison, anyons in fractional quantum Hall systems systems may satisfy properties (i) and (ii), but  are not protected from thermal fluctuations. Majoranas in topological superconductor nanowires do display all properties above, but offer no path towards universal gates by braiding. The Kondo approach we put forward displays all qualities discussed above. Here, the 3-channel Kondo case will be pivotal, given the suggestive signatures of it displaying Fibonacci anyons -- which \emph{do} braid to universal gates.
 ~Realizing Fibonacci particles is a daunting task; all proposals we are aware of require challenging fine-tuning and/or hard-to-realize ingredients.~\cite{Read-Rezayi,Mong_coupled,Hu_coupled,Lopes_coupled} In fact, even the engineering of Majorana zero-modes in nanowires presents a challenging goal. In contrast, 2-channel and 3-channel Kondo phenomenology has already been reported in devices based on charge implementations of the impurity pseudospin.~\cite{Iftikhar_3CK}

Our work lays the foundation for the implementation of quantum computation in Kondo systems. Yet, several important extensions of this work can be envisioned. As noteworthy topics we mention a more detailed modelling for a device realization, a better microscopic description of the multi-fusion ansatz, and establishing a relation of the two-point correlation functions to accessible observables such as conductance or tunneling measurements. For these latter two points, we have been exploring large-number-of-channels perturbative calculations which, preliminarily, are in agreement with the results here described.~\cite{LargeK} This limit provides interesting insight, as the spins are perturbatively free from the conduction channels and play the roles of the free anyonic modes themselves. The measurement of fusion-channel-fixed correlations is a more subtle issue. If possible, we still note that such measurements can be challenging in some scenarios, demanding interferometry. An example is the 2-channel Ising case, whose correlation functions were demonstrated to change only by a phase, depending on the fusion channel. Remarkably, we saw that the most interesting and useful Fibonacci 3-channel case involves a suppression of the correlation function in modulus, not just in phase, alleviating the requirement of interferometry. In a recent proposal for Fibonacci anyons, a conductance that depends on the fusion outcome exactly as given by Eq.~\eqref{eq:effcorr} was proposed.~\cite{Hu_coupled}

The realization of platforms for topological quantum computation is a grand challenge from the point of view of condensed matter physics, quantum phases and quantum information. By relying on frustration, strong correlation and localization effects, multi-channel Kondo systems function in a regime totally different from gapped topological order. Still, they display several of the desired properties of standard anyon systems. This paradigm shift presents several advantages but, most importantly, surmountable and exciting challenges.

\section{Acknowledgments}
We would like to thank J. Folk, A. Nocera, K. Shtengel, R. Boyack and S. Plugge for comments and discussions, as well as Y. Oreg and  H. S. Sim for insightful discussions motivating this topic. This work has been supported by the Canada First Research Excellence Fund, the NSERC Discovery Grant 04033-2016, and by the US-Israel Binational Science Foundation (Grant No. 2016255) (E.S.).

\bibliographystyle{apsrev4-1}
\bibliography{main.bbl}

\appendix

\section{Chiral device toy-model \label{sec:dev_toy}}

To study the spin asymmetry effects in the proposed device, we focus on a single effective impurity and a single channel and consider a simplified Anderson-impurity phenomenological model $H=H_0+H_1$, where
\begin{align}
    H_{0}&=U\left(\sum_{\sigma}f_{\sigma}^{\dagger}f_{\sigma}-1\right)^{2}+\left(\sum_{\sigma}\epsilon_{\sigma}f_{\sigma}^{\dagger}f_{\sigma}-\left(\epsilon_{\uparrow}+\epsilon_{\downarrow}\right)/2\right)\\H_{1}&=\left(\sum_{\sigma}t_{\sigma}f_{\sigma}^{\dagger}c_{\sigma,0}+H.c.\right).
\end{align}
This model has close similarities to that of Ref.~\onlinecite{Kikoin_oreg_2CK}. The dispersion of $c$ fermions is unimportant in comparison with other energy scales, so we focus only on the impurity site. Under this consideration, we discretized space in a lattice, with $c_{\sigma,0}$ corresponding to an electron at the origin with spin $\sigma$. The two energy levels of the QD are being modelled by a fermionic representation of associated Hubbard operators given by $f_\sigma$;~\cite{Kikoin_oreg_2CK} in practice, states $\left|\pm\right\rangle =f_{\pm}^{\dagger}\left|0\right\rangle $ correspond to the triplet and singlet states forming the lowest two energy levels of the dot. The interaction energy $U$ 
removes the other triplet states of the problem, which correspond to the doubly and unoccupied levels of fermions $f$. We ignored the Zeeman splitting for the $c_\sigma$ electrons, but consider its consequences as a spin-dependent tunneling amplitude $t_\sigma$. To be general, however, we do need to include an imperfect accidental degeneracy of the relevant impurity energy levels $\epsilon_{\sigma}$. The particle-hole transformation, as implemented by $c_{\sigma,0}\to c_{\sigma,0}^{\dagger}$  and $f_{\sigma}\to-f_{\sigma}^{\dagger}$, preserves this Hamiltonian but for a flip $\epsilon_{\sigma}\to-\epsilon_{\sigma}$. In the particle-hole symmetric limit, we have single occupation of both fermions: $\sum_{\sigma}c_{\sigma,0}^{\dagger}c_{\sigma,0}=\sum_{\sigma}f_{\sigma}^{\dagger}f_{\sigma}=1$.

At large $U$, an effective Hamiltonian can be computed by a Schrieffer-Wolff transformation.~\cite{Schrieffer_Wolff_Kondo} We leave the details for Appendix~\ref{sec:SW_app}, to preserve the flow of the discussion. Neglecting fluctuations of particle number, the effective Hamiltonian reads 
\begin{align}
    H&\approx H_{0}+H_{2} \nonumber \\&=U\left(n_{f}-1\right)^{2}+\epsilon_{0}\left(n_{f}-1\right) \nonumber \\&-B_{0}\mathcal{S}_{z}+\left(2\epsilon_{z}+B_{0}\right)S_{z} \nonumber \\&+g_{z}\mathcal{S}_{z}S_{z}+g_{\perp}\left(\mathcal{S}_{x}S_{x}+\mathcal{S}_{y}S_{y}\right). \label{eff_H}
\end{align}
As in the main text, $g_\perp=g$ and $g_z=\Delta g$. Here, $\epsilon_{\sigma}=\epsilon_{0}+\sigma\epsilon_{z}$ and $\epsilon_z$ corresponds to the unrenormalized Zeeman splitting of the dot singlet and triplet; then
\begin{align}
    \boldsymbol{\mathcal{S}}=\frac{1}{2}c_{0}^{\dagger}\boldsymbol{\sigma}c_{0},   ~~~~  \boldsymbol{S}=\frac{1}{2}f^{\dagger}\boldsymbol{\sigma}f,
\end{align}
and
\begin{align}
    B_{0}=&\sum_{\sigma}\frac{\sigma t_{\sigma}^{2}\epsilon_{-\sigma}}{\left(\epsilon_{-\sigma}+U\right)\left(\epsilon_{-\sigma}-U\right)}\\g_{z}=&2\sum_{\sigma}\frac{Ut_{\sigma}^{2}}{\left(U+\epsilon_{-\sigma}\right)\left(U-\epsilon_{-\sigma}\right)}\\g_{\perp}=&2\sum_{\sigma}\frac{Ut_{\uparrow}t_{\downarrow}}{\left(U+\epsilon_{-\sigma}\right)\left(U-\epsilon_{-\sigma}\right)}.
\end{align}
Two points call our attention in these equations. First, the previously alluded spin asymmetric couplings between the two spinful channels and the QD are explicitly manifest. As mentioned, a spin-anisotropy is not expected to affect the NFL low-energy physics. More worrying is the effective magnetic field term $B_0$, which threatens to break the bound-state that overscreens the impurity at the origin. From a field theory perspective, however, the magnetic field coupling to the impurity dominates over the coupling to the conduction electrons.~\cite{Barzykin_PRB} Fortunately, this term may be tuned via $B_{\parallel}$ to zero  under the condition of degeneracy of the two renormalized QD levels. Setting $B_0=-2\epsilon_z$ leads to, in the large $U$ limit,
\begin{equation}
    \epsilon_{z}\approx\frac{\left(t_{\uparrow}^{2}-t_{\downarrow}^{2}\right)\epsilon_{0}}{2U^{2}}.
\end{equation}
For equal hybridization amplitudes $t_\uparrow=t_\downarrow$, there is no spin-asymmetry or splitting renormalization; the accidental degeneracy must be tuned exactly to guarantee the absence of the effective magnetic field. If the hybridization amplitudes are spin-dependent, however, we may still tune the effective magnetic field for the impurities away by the real applied magnetic field which controls $\epsilon_z$. Note also that even in the extreme asymmetric situations, when one of the $t_\sigma$ is set to zero, the Kondo couplings are well-established, showing some robustness in our proposal.

\section{Schrieffer-Wolf transformation \label{sec:SW_app}}
Defining the generator $S$ satisfying $\left[H_{0},S\right]=H_{1}$, the correction to the Hamiltonian is given by $H_{2}=\frac{1}{2}\left[S,H_{1}\right]$. The generator reads
\begin{equation}
    S=\sum_{\sigma,\alpha}\frac{t_{\sigma}}{\Delta_{\sigma}^{\alpha}}n_{-\sigma}^{\alpha}c_{\sigma,0}^{\dagger}f_{\sigma}-H.c.,
\end{equation}
where $\Delta_{\sigma}^{\alpha}=\left(\left(1-\sigma\right)\epsilon_{\uparrow}+\left(1+\sigma\right)\epsilon_{\downarrow}\right)/2+\alpha U$, for $\alpha=\pm$, are the energy differences for double to single and single to zero occupations. The terms $n_{-\sigma}^{\alpha}=\text{sgn}\left(\alpha\right)\left(n_{-\sigma}\right)+\theta\left(-\alpha\right)$, where $n_{\sigma}=f_{\sigma}^{\dagger}f_{\sigma}$, act as projector operators.~\cite{Schrieffer_Wolff_Kondo} We compute the effective Hamiltonian under the approximation of substituting $\sum_{\sigma}c_{\sigma,0}^{\dagger}c_{\sigma,0}=\sum_{\sigma}f_{\sigma}^{\dagger}f_{\sigma}=1$ (and consequently $n_{+}n_{-}\to0$). This neglects, to lowest order, particle density fluctuations at the impurity site. If we recovered the particle-density fluctuations, the effective magnetic field terms in $H_2$ would contain cross terms between spin and particle densities fluctuations around the single occupation mean value. Since we find that $B_0$ can tune the impurity splitting to zero anyway, and that which is the most important perturbation,~\cite{Barzykin_PRB} we consider the lowest approximation in fluctuations for simplicity of analysis. The result we find is displayed in Eq.~\ref{eff_H} in the main text.

The Hamiltonian we find is slightly different from that of Ref.~\onlinecite{Kikoin_oreg_2CK}, although the models are closely related. While part of the difference arises from us taking the zero density fluctuation limit -- which we stress is not necessary -- we do found some other differences. We found, in particular, effective local magnetic field contributions both to the impurity and to the conduction electrons, independently of the approximation. We believe this discrepancy stems from a careful normal ordering and anti-commutation bookkeeping during Schrieffer-Wolff/second-order perturbation theory calculation.

\section{Scalar scattering Green's functions \label{sec:sca_green}}
The 2-point correlation (Green's) functions can be obtained by elementary methods. In a finite-size and finite-temperature torus, it reads
\begin{align}
    &G^{\mathrm{torus}}\left(\tau,x,\tau',x'\right)\nonumber \\&=\frac{1}{L\beta}e^{i2\pi\sum_{l=1}^{M}V^{\left(l\right)}\left[\theta\left(x-x_{l}\right)-\theta\left(x'-x_{l}\right)-\frac{x-x'}{L}\right]} \nonumber \\&\times\sum_{m,n}\frac{e^{i\frac{2\pi}{L}\left(m+\frac{1}{2}\right)\left(x-x'\right)}e^{-i\frac{2\pi}{\beta}\left(n+\frac{1}{2}\right)\left(\tau-\tau'\right)}}{i\omega_{n}-k_{m}}.
\end{align}
It is instructive to consider this expression both in the thermodynamic and zero-temperature limits, in which cases the sums can be performed exactly. In the former case, we have
\begin{align}
    &\lim_{L\to\infty}G^{{\rm{torus}}}\left(\tau,x,\tau',x'\right) \nonumber \\=&\frac{\pi}{\beta}\frac{e^{i2\pi\sum_{l=1}^{M}V^{\left(l\right)}\left[\theta\left(x-x_{l}\right)-\theta\left(x'-x_{l}\right)\right]}}{\sin\frac{\pi}{\beta}\left(z-z'\right)},
\end{align}
while in the latter
\begin{align}
    &\lim_{T\to0}G^{{\rm{torus}}}\left(\tau,x,\tau',x'\right) \nonumber \\&=\frac{\pi}{L}\frac{e^{i2\pi\sum_{l=1}^{M}V^{\left(l\right)}\left[\theta\left(x-x_{l}\right)-\theta\left(x'-x_{l}\right)\right]}e^{\frac{2\pi}{L}\left(\sum_{i=1}^{M}V^{\left(l\right)}\right)\left(z-z'\right)}}{\sinh\frac{\pi}{L}\left(z-z'\right)}.
\end{align} 
 It is known in the context of the NCK problem that the anyonic behavior of the overscreened impurity only develops in the thermodynamic limit, followed by a low temperature limit. The order of the limits is important.~\cite{Affleck_Kondo_review,Rozhkov_imp_entropy} Some phenomenon reminiscent of that also appears  here. In the expressions above, one sees already that in the $L\to \infty$ limit, the behavior of the correlation function is that of absorbing a simple prefactor correction that changes discontinuously with the number of impurities between the points $x$ and $x'$. Taking the zero-temperature limit first leaves us with an expression that remembers that the system is defined in a cylinder and has to display the periodicity-correcting factors which arise due to the momentum shifts.

The scalar scattering problem, however, does not display impurity dynamics and interaction effects. As a consequence, the order of limits discussed above is less stringent when one finally computes the correlation function in the plane (both $L\to\infty$ and $T\to0$). Independently of the order of limits, we obtain the result displayed in the main text:
\begin{equation}
    G^{\mathrm{plane}}\left(\tau,x,\tau',x'\right)=\frac{e^{i2\pi\sum_{l=1}^{M}V^{\left(l\right)}\left[\theta\left(x-x_{l}\right)-\theta\left(x'-x_{l}\right)\right]}}{z-z'}.
\end{equation}
As discussed in the main text, this result could be have been found directly computing the zero-temperature correlation function from Eq.~\eqref{eq:contPSI}.

\section{2CK eigenmodes and multi-impurity absorption \label{sec:2CK_eig_app}}
Here we display some extra details on the analysis of the 2CK problem with multiple impurities. Namely, our goal is to prove that the Majorana modes are all absorbed by the conduction electrons at each impurity site, and consider the associated phase shifts. Similar results are effective in the single-impurity case, and we show here that they also generalize for many impurities.

To start, we decompose the field operators according to~\cite{Sela1,Sela2} (we drop the labels, from now on only $\chi_{sf}^1\to \chi$ is part of the discussion)
\begin{align}
    \chi\left(x\right)&=\sum_{k}\varphi_{k}\left(x\right)c_{k},~~~b_{l}=\sum_{k}u_{k}^{l}c_{k}.
\end{align}
The wavefunctions satisfy the Schr\"odinger equations
\begin{align}
    i\partial_{x}\varphi_{k}\left(x\right)+i g_{\perp}\sum_{l}2\pi\delta\left(x-x_{l}\right)u_{k}^{l}&=k\varphi_{k}\left(x\right)\\i g_{\perp}\varphi_{k}\left(x_{l}\right)&=ku_{k}^{l},
\end{align}
under the local constraints $\varphi_{k,l}^{\left(-\right)}=e^{i2\delta}\varphi_{k,l}^{\left(+\right)}$, where $\varphi_{k,l}^{\left(\pm\right)}\equiv\lim_{\epsilon\to0}\varphi_{k}\left(x_{l}\pm\epsilon\right)$ and
\begin{equation}
    \tan\delta=\frac{2\pi g_{\perp}^{2}}{k}.
\end{equation}
In the decoupled limit, $\delta\to0$, while in the strong coupling limit $\delta\to\pi/2$. The latter happens as $g_{\perp}^{2}\sim T_{K}\gg\Lambda$, where $\Lambda$ is the characteristic energy scale of the free modes. Carefully sewing the solutions through the impurities results in
\begin{align}
    &\varphi_{k}\left(x\right)=e^{-ik\left(x-x_{1}\right)}\varphi_{k,1}^{\left(-\right)}\nonumber \\&\times\left[1+\left(1-e^{i2\delta}\right)\sum_{l=1}^{M}\theta\left(x-x_{l}\right)e^{-i2l\delta}\right].
\end{align}
In the strong coupling limit, $\delta=\pi/2$ and the conduction-electron field operators are excluded from the impurity positions -- as the term in brackets guarantees. At those points, the impurity Majoranas take over, complementing and completing the states. To evince this, we define
\begin{align}
    \tilde{\chi}\left(x\right)&\equiv\chi\left(x\right)\left[1+2\sum_{l=1}^{M}\left(-1\right)^{l}\theta\left(x-x_{l}\right)\right] \nonumber \\&=\left(-1\right)^{l+1}\sum_{k}e^{-ik\left(x-x_{l}\right)}\varphi_{k,l}^{\left(-\right)}c_{k},
\end{align}
which is finite at the impurity sites in the strong coupling limit. In fact, by the impurity Majorana field operator expansion, we have the operatorial identity
\begin{equation}
    b_{l}\approx\frac{\left(-1\right)^{l}}{\pi g_{\perp}}\tilde{\chi}\left(x_{l}\right),
\end{equation}
proving that this identity, which was true in the single impurity case,~\cite{Sela1,Sela2} also remains true here. Finally, note the phase build-up
\begin{align}
    \chi\left(L\to\infty\right)&=e^{-i2M\delta}e^{-i2kL}\chi\left(-L\to\infty\right) \nonumber \\&\underset{\delta\to\pi/2}{\to}e^{-iM\pi}e^{-i2kL}\chi\left(-L\to\infty\right).
\end{align}
For an even number of impurities, no phase is gathered (except from the one gained from translation), while for an odd number of impurities one always picks up an overall $\pi$ total phase. Notice, that the spin-flavor Majorana operators are not the original physical degrees of freedom. They are, instead, highly correlated collective modes, not directly related to any obvious measurement (such as the tunneling currents of our focus). Yet, this boundary condition flip has useful consequences when discussing correlations.~\cite{MALDACENA_Ludwig_SO8}

\end{document}